\documentclass[
showpacs,
amsmath,
amssymb,
aps,
reprint]{revtex4-1}

\usepackage[utf8]{inputenc}
\usepackage[english]{babel}
\usepackage{esint}
\usepackage{braket}
\usepackage{dcolumn}
\usepackage{verbatim}

\usepackage{graphicx}
\usepackage{epstopdf} 
\usepackage[dvipsnames]{xcolor}
\usepackage[normalem]{ulem}
\usepackage{hyperref}
\usepackage{slashed}
\usepackage{dsfont}
\usepackage{float}
\usepackage{tikz}
\usetikzlibrary{snakes}
\usepackage{latexsym}
\usepackage{epsf}

\newcommand{\be}[1]{ \begin{eqnarray} \mbox{$\label{#1}$} }
   
\newcommand{\ee}{\end{eqnarray}}

\newcommand{\eeq}{\end{equation}}

\newcommand\ie {{\it i.e. }}
\newcommand\eg {{\it e.g. }}

\newcommand\half{\frac 1 2 }

\newcommand{\etal} {{\em et al. }}
  \DeclareMathOperator{\g}{\gamma}
  \DeclareMathOperator{\Id}{\mathbb{I}}
  \DeclareMathOperator{\Tr}{Tr}
  \DeclareMathOperator{\sgn}{sgn}

\newcommand\noi{\noindent}

\newcommand\oncite [1] {Ref. \onlinecite{#1} }
\newcommand\oncitep [1] {Ref. \onlinecite{#1}. }
\newcommand\oncitec [1] {Ref. \onlinecite{#1}, }

\def\g{\gamma}

 \newcommand{\mbk} {\mathbf{k}}

\newcommand\emn{\epsilon^{\mu\nu\sigma\lambda}}

\begin{document}
\title{On the Electromagnetic Response of Topological Superconductors}

\author{M. St\aa lhammar$^1$}
\author{M. Stone$^2$}
\author{Masatoshi Sato$^3$}
\author{T.H. Hansson$^1$}

\affiliation{$^1$Department of Physics, Stockholm University, AlbaNova University Center, 106 91 Stockholm, Sweden}
\affiliation{$^2$Department of Physics, University of Illinois, 1110 West Green Street, Urbana, Illinois 61801, USA}
\affiliation{$^3$Yukawa institute for Theoretical Physics, Kyoto University, 606-8502, Japan}

\begin{abstract}
We resolve several puzzles related to the electromagnetic response of topological superconductors in 3+1 dimensions. In particular we show by an analytical calculation that the interface between a topological and normal superconductor does not exhibit any quantum Hall effect as long as time reversal invariance is preserved. We contrast this with the analogous case of a topological insulator to normal insulator interface. The difference is  that in the topological insulator  the electromagnetic vector potential couples to a vector current in a theory with a Dirac mass, while  in the superconductor   a pair of   Weyl fermions are  gapped by  Majorana masses  and the electromagnetic vector potential couples to their axial currents.

\end{abstract}
                                        
\maketitle                 

\section{Introduction}                  

The ten-fold way classification of  topological matter, and in particular topological insulators (TI), and topological superconductors (TSC) was originally derived using non-interacting fermions \cite{PhysRevB.78.195125,kitaev2009}.  By studying the anomaly structure it was later understood how this classification generalizes to interacting systems \cite{PhysRevB.85.045104}. A principal tool in this analysis is the notion of an \emph{effective action} that governs the response of the system to external probe fields. For topological insulators the natural probe is electromagnetic, but this appears to be problematic for superconductors since the Mei\ss ner effect confines the electromagnetic fields to boundaries or vortex cores. However, because topological effects \emph{are} often related to surfaces or defects,    characteristic features of TSCs may yet be coded into an effective topological field theory (TFT) for the electromagnetic field. 

Such a  TFT  was  indeed  proposed some time ago by Qi, Witten and Zhang \cite{qwz},  who argued that the action governing the  low energy electromagnetic response  of  a TSC contains  a topological  term $\sim  \int \theta F\tilde F$, where $F $ is the electromagnetic field tensor and $\tilde F$ its dual.  Although in appearance this term is similar to the $\theta$-term in the effective action for a TI, it differers in that, rather than  being an external parameter that distinguishes the trivial from the non-trivial phase,  $\theta $ is a \emph{dynamical} field, namely the phase of the SC order parameter. Despite this difference,  Qi \etal argued  that the presence of this term implies  topological effects which could, at least in principle, be observable. 

Later, in  \oncitec{sl} Stone and Lopes raised several questions about the claims made by Qi \etal  First they recalculated the effective action, and found a coefficient in front of the $\theta$-term that differed from that in  \oncitec{qwz} and secondly they pointed out a problem with  the  Qi \etal prediction  that the  $\theta$-term  implies  that  an applied electric field, in addition to the expected Josephson-type longitudinal acceleration of  the superfluid,  drives a {\it transverse} current into a vortex core when the   field is parallel to the vortex. In   \oncite{qwz} it was suggested that this inflow would be absorbed by an anomaly in the one-dimensional (1D) gapless mode in the vortex core. However  the core-bound  mode is a  neutral \emph{Majorana} mode and is therefore incapable of absorbing the charge inflow.

A related issue  occurs for a system with a boundary.  Here the $\theta$-term  suggests that a TSC displays a surface Hall conductance similar to that carried by the charged gapless surface mode  of a 3D  TI.  Related to this one also expects a charge accumulation where a Abrikosov vortex crosses the interface, similar to the one predicted for the TI case in \oncitep{PhysRevLett.117.167002} Again, however, the TSC surface exhibits a {\it neutral}  2D Majorana mode which appears incapable of supporting a Hall current. These two problems are clearly  related as we could model a vortex in a TSC as a cylinder of  ordinary superconductor  containing an Abrikosov vortex, in which case the putative charge transport would occur  at the surface of the cylinder.  

In this paper we revisit these  puzzles. 
By a direct perturbative calculation we derive the gauge invariant topological current for the infinite TSC and also the Chern-Simons (CS) term that determines the low-energy properties of a planar interface between a TSC and a normal SC. For the current we obtain the same result as in \oncite{sl} and in particular the same coefficient for the  $\theta$-term.  For the CS term we use the  techniques developed by Mulligan and Burnell in \oncite{mb} for the case of a TI-I interface  to calculate the CS coefficient  as a function of the  time reversal  invariance (TRI)  breaking term.  We show that when the symmetry breaking is such that the $\theta$-angle winds slowly compared to the mass gap then  the CS coefficient coincides with  that predicted by total change in the $\theta$-angle  in  \oncitec{sl} but as as the TRI breaking term is turned off and $\theta$ winds rapidly then  the the CS coefficient becomes smaller and vanishes in the limit of vanishing TRI breaking. We contrast this with  the behavior at the TI-I interface investigated in  \oncite{mb}, where the CS coefficient remains finite as the TRI breaking term is removed. 

We are thus able to reconcile the apparent conflict of the  existence of the of a $\theta$-term in the low-energy effective theory for the TSC with the fact that there is  no CS term, and thus no Hall effect,  at  a TSC-SC interface in the limit where TRI is preserved. The  essential difference between the two cases are that in the effective low-energy action for the TSC, the electromagnetic potential couple as an axial gauge field to Majorana fermions gapped by a Majorana mass, rather than as a vector field to Dirac fermions with a Dirac mass. 

The paper is organized as follows. In Section \ref{sec:effact} we first show by a direct calculation how the Majorana theory with an axial coupling to the electromagnetic potential, which was written in  \oncite{sl} based on symmetry considerations, directly follows from a mean field  Hamiltonian and then how to obtain the topological current, and the response action that depends on the electromagnetic potential, and the SC phase fields.  In Section \ref{sec:cstheory}, which contains our main results, we calculate the Hall response at a TSC-SC boundary, discuss its topological significance and  show that it vanishes as the TRI breaking term is removed. Finally, in Section \ref{sec:res/sum}, we discuss  the effects of screening that are always present in real superconductors, and which are important to include in order to resolve an apparent contradiction. We conclude Section \ref{sec:res/sum} with a short summary of our results.

%%%%%%%%%%%%%%%%%%%%%%%%%%%%%%%%%%%%%%%%%%%%%%%%
%%%%%%%%%%%%%%%%%%%%%%%%%%%%%%%%%%%%%%%%%%%%%%%%

\section{Elementary derivation of the effective response action} \label{sec:effact} 

 First we derive the appropriate low-energy fermion theory directly from a microscopic BdG Hamiltonian, and in subsection \ref{sub:cstheory} we analyze its topological properties, especially with respect to anomalies. Finally in  \ref{sub:efftheory} we give the pertinent low-energy theory for the TSC to be used later.

%%%%%%%%%%%%%%%%%%%%%%%%%%%%%%%
\subsection{From Hamiltonian to Dirac action}  \label{sub:hamiltonian}

We start from the two band Hamiltonian used in  \oncite{qwz} by  Qi \etal 
\be{ham}
H = \half \sum_\mbk (c_\mbk^\dagger, c_{-\mbk} )  \begin{pmatrix}
h_\mbk & \Delta_\mbk \\ \Delta^\dagger_\mbk & - h^\star_{-\mbk}  \end{pmatrix}
\begin{pmatrix} c_\mbk \\ c^\dagger_{-\mbk}
\end{pmatrix}\ee
where we refer to the band index as spin, and use the notation  $c_\mbk = (c_{\mbk\uparrow} , c_{\mbk\downarrow} )^T$.  The kinetic energy operator is 
$
h_\mbk = \frac {k^2} {2m} - \mu   + \alpha \vec\sigma\cdot\vec k
$ and  the pairing function is $\Delta_\mbk =-i\Delta \vec\sigma\cdot\vec k \sigma_y$, so
$
\Delta^\dagger_\mbk = i \Delta^\star \sigma_y \vec\sigma\cdot\vec k 
$. 
Introducing the Nambu 4-spinor,
\be{spinor}
\Psi_\mbk =  \begin{pmatrix}   c_\mbk \\ -i\sigma_y c^\dagger_{-\mbk}
\end{pmatrix}
\ee
we can rewrite Eq.~\eqref{ham} as
\be{ham2}
H = \half  \sum_\mbk \Psi_\mbk^\dagger  \begin{pmatrix} h_\mbk & \Delta\, \vec\sigma\cdot\vec k \\ 
 \Delta^\star\, \vec\sigma\cdot\vec k  & -h_{-\mbk} \end{pmatrix} \Psi_\mbk \, .
\ee
Next we define the helicity eigenstates,
\be{helicity}
\vec\sigma\cdot\vec k \ket{\mbk,\pm} = \pm k \ket{\mbk,\pm}
\ee
with $E=0$ defining the Fermi wave vectors for the helicities $k_{F,\pm} =\sqrt {m^2\alpha^2 + 2m\mu}$. In the following we shall
put the chiral symmetry breaking parameter $\alpha$ to zero and thus have a common Fermi wave vector $k_F$. 
%We take equal chemical potential for the two chiral Fermi sursfaces, $\mu_+ = \mu_- = \mu$

The  low-energy spectrum is obtained by expanding $\vec k \rightarrow (k_F + k) \hat k$ to linear order in $\vec k$.
 and the low-energy excitations are determined by,
\be{lowe}
h_\mbk   = \pm v_F \vec\sigma \cdot \vec k + \mathcal{O}(k^2).
\ee
Switching to the helical basis $\Psi_{\mbk \pm}$  and doing some algebra gives, 
\be{hamfin}
H = \pm \half \sum_\mbk \Psi^\dagger_{\mbk \pm}  \begin{pmatrix}
h_\mbk &  k_F \Delta\\ k_F \Delta & \mp h^\star_{-\mbk}  \end{pmatrix} \Psi_{\mbk \pm}.
\ee
Using the notation   $\Delta k_F \rightarrow |\Delta| e^{i\theta}$ and  the Weyl gamma matrices,
\be{weyl}
\gamma^0 = \begin{pmatrix} 0 & \mathds{1} \\ \mathds{1}  & 0\end{pmatrix};\quad
\gamma^5 = \begin{pmatrix} - \mathds{1} & 0 \\ 0 &\mathds{1}  \end{pmatrix} ;\quad
\vec\gamma = \begin{pmatrix} -  \vec\sigma & 0 \\ 0 & -\vec\sigma \end{pmatrix}
\ee
Heisenberg's equations of motion and some more algebra gives,
\be{eom}
\left( i\gamma^0\partial_0 \mp i\gamma\cdot\vec k \mp \Delta e^{i\theta_\pm\gamma^5} \right) \Psi_\pm = 0 \, .
\ee
where, because of \eqref{spinor},  $\Psi_\pm$ are Majorana spinors.

To write \eqref{eom} in a unified form in coordinate space, we redefine the left-handed spinor as $\Psi_- \rightarrow \gamma^0 \Psi_-$ to finally get,
\be{findirac}
\left( i\slashed\partial - \Delta\, e^{\pm i\theta_\pm\gamma^5} \right) \Psi_\pm = 0 \, ,
\ee
where $\Delta$ is understood to be real. The electromagnetic gauge transformation $c_\mbk \rightarrow e^{-i\lambda} c_\mbk$  acts  on $\Psi_{\mbk \pm}$  as, 
\be{emgt}
\Psi_\pm \rightarrow e^{\pm i \gamma^5 \lambda}  \Psi_\pm;\quad \theta_\pm \rightarrow \theta_\pm \mp 2 \lambda.
\ee
To restore local gauge invariance we couple to the gauge field as,
\be{gifindirac}
\left( i\slashed\partial \pm \slashed A \gamma^5 - \Delta e^{\pm i\theta_\pm\gamma^5} \right) \Psi_\pm = 0 \, ,
\ee
where $A$ transforms as $A_\mu \rightarrow A_\mu + \partial_\mu \lambda$. Note that the electromagnetic potential 
couples to the \emph{axial} current. This is typical of superconductors, and is consistent with  particle number   not being conserved in the presence of a pairing term, which here
appears as a mass. In fact $\partial_\mu j_5^\mu \equiv  \partial_\mu j^\mu_{\text{em}} \sim \Delta \ne 0$, and particle number  conservation is only restored when $\Delta$ is allowed to obey  its own equation of motion.

The action corresponding to Eq.~\eqref{gifindirac}, 
\be{lagdirac}
\mathcal{L} =  \bar \Psi_\pm  \left( i\slashed\partial \pm \slashed A \gamma^5 - \Delta e^{\pm i\theta_\pm\gamma^5} \right) \Psi_\pm
\ee
was postulated earlier in \oncite{sl} based on symmetry arguments. It is important that the requirement of TRI implies that the angles $\theta_\pm$ can only take the values $\pm \pi$. Superficially we have a description similar to that of a TI, and would thus expect that integrating out the fermions would give a topologial  $\theta$-term, $\theta/(32\pi^2)  \int F \tilde F$, in the response action, which differs from the TI result by a factor of two because of the Majorana condition implied by Eq.~\eqref{spinor}. This is the result given in \oncite{qwz}.

%%%%%%%%%%%%%%%%%%%%%%%%%%%%%%%
\subsection{Topological currents and actions}  \label{sub:cstheory}

In calculating the effective low-energy action we shall concentrate on the topological, or anomalous, parts that involve the Levi-Civita symbol. There is also  a non-topological part that describes the usual supercurrent and  encodes the Mei\ss ner effect, but for the discussion in this section we do not need to specify but just keep in mind that it is gauge invariant, and thus a function of the covariant derivatives $\partial_\mu \theta_\pm - 2e A_\mu$.
In a superconductor the  angles $\theta_\pm$ are the phase fields   $\theta_\pm$ of  the superconducting order parameter gapping  the  left $(+)$ and right-handed  $(-)$ Fermi surfaces, and as such they are dynamical. This is a key distinction between topological superconductors and topological insulators, where the $\theta$-angles are fixed  in the bulk. In  both cases, however, the theory preserves TRI only if  the difference $\theta_+-\theta_-$ is restricted to equal 0 or $\pi$. However, before coming to the effective action we show how to extract the topological part of the current with a minimum of assumptions.

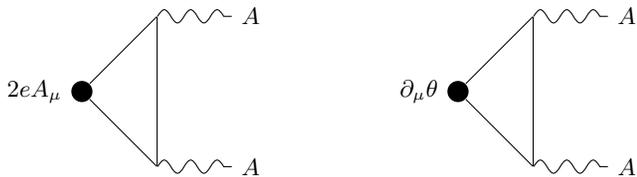
\begin{figure}
\begin{center}
\begin{tikzpicture}[shorten >=1pt,->]
  \tikzstyle{vertex}=[circle,fill=black!100,minimum size=8pt,inner sep=0pt]
  \tikzstyle{vertex1}=[circle,fill=black!100,minimum size=0pt,inner sep=0pt]
  \node[vertex][label=left: $2eA_{\mu}$]  (G_11) at (-3,-1) {} ;
  \node[vertex1] (G_12) at (-2,0)   {};
  \node[vertex1] (G_13) at (-2,-2)  {};
  \node[vertex1][label=right: $A$] (G_14) at (-1,-2) {} ;
  \node[vertex1][label=right: $A$] (G_15) at (-1,0) {} ;
  \draw (G_11) -- (G_12) -- (G_13) -- cycle ;
  \draw (G_11) -- (G_13) -- cycle;
  \draw[snake=coil,segment aspect=0] (G_13) -- (G_14) --cycle ; 
  \draw[snake=coil,segment aspect=0] (G_12) -- (G_15) --cycle ; 
  
  \tikzstyle{vertex}=[circle,fill=black!100,minimum size=8pt,inner sep=0pt]
  \tikzstyle{vertex1}=[circle,fill=black!100,minimum size=0pt,inner sep=0pt]
  \node[vertex][label=left: $\partial_{\mu}\theta$]  (G_1) at (2,-1) {} ;
  \node[vertex1] (G_2) at (3,0)   {};
  \node[vertex1] (G_3) at (3,-2)  {};
  \node[vertex1][label=right: $A$] (G_4) at (4,-2) {} ;
  \node[vertex1][label=right: $A$] (G_5) at (4,0) {} ;
  \draw (G_1) -- (G_2) -- (G_3) -- cycle ;
  \draw (G_1) -- (G_3) -- cycle;
  \draw[snake=coil,segment aspect=0] (G_3) -- (G_4) --cycle ; 
  \draw[snake=coil,segment aspect=0] (G_2) -- (G_5) --cycle ; 
\end{tikzpicture}
\end{center}
\caption{Diagrammatic illustration of the process resulting in the current Eq.~\eqref{aatheta_covariant}.}\label{fig:diag}
\end{figure}

 We begin with  the currents that can be determined from the diagrams in Figure \ref{fig:diag}. In \oncite{sl} it was stressed that the $AA\partial\theta$ diagram is convergent and a left or right-handed Weyl spinor unambiguously contributes 
\be{aatheta}
j_{\theta;\pm}^\mu   = \pm  \frac e {48\pi^2} \emn \partial_\nu\theta_\pm F_{\sigma\lambda}
\ee
to the axial current. (Recall that for our  superconductor  the ``axial''  current is   the physical electromagnetic  current.) 

On the other hand, the  $AAA$ triangle graph in Figure \ref{fig:diag} is only {\it conditionally\/} convergent, and the  result depends on how the momenta are routed (a clear exposition stressing this point is given in \oncite{srednickibook}). There are two ways to select   its value. The first is to demand that the topological  current is  invariant under the gauge transformation,
 \be{EQ:gauge_transformation}
 \theta_\pm &\to& \theta_\pm -2\alpha e,\nonumber\\
 A_\mu &\to& A_\mu +\partial_\mu \alpha,
 \ee
and for this to be true, the momentum in the $AAA$ diagram must be routed so to  augment the $\partial_\mu  \theta$  in Eq.~\eqref{aatheta} to a gauge-covariant derivative. The resulting currents thus become, 
\be{aatheta_covariant}
j_{\pm}^\mu   = \pm  \frac e {48\pi^2} \emn( \partial_\nu\theta_\pm + 2e A_\nu )F_{\sigma\lambda} \, .
\ee

To prove that the expression in Eq.~\eqref{aatheta_covariant} for the topological current, which was derived using perturbation theory and demanding gauge invariance, is indeed an exact result, we use a second way to determine how to define the $AAA$ graph. Closely following  \oncitec{sl} we show how it follows from an effective action  which is tightly constrained by the anomaly structure. First, note that the piece in the current in Eq.~\eqref{aatheta_covariant} which involves $\partial_\nu \theta$ can be obtained from an action as,
\be{constcurr}
j_{\theta;\pm}^\mu= -\frac{\delta S_{\theta; \pm}}{\delta A_\mu(x)} 
\ee
where 
\be{4top}
S_{\theta; \pm} =   \mp  \frac e {192\pi^2} \int d^4x \, \theta_\pm \emn F_{\mu\nu} F_{\sigma\lambda}  \, ,
\ee
and that neither the action, nor the corresponding current, Eq.~\eqref{constcurr}, is gauge invariant.
The action Eq.~\eqref{4top}, which can also be obtained directly by making a chiral rotation of the original theory and using  the (finite) form of the  \emph{consistent} chiral anomaly, 
 indeed transforms as
\be{}
\delta_{\alpha} S_{\theta; \pm}=\pm \alpha\, \frac e {96\pi^2} \emn F_{\mu\nu} F_{\sigma\lambda} \, ,
\ee
under Eq.~\eqref{EQ:gauge_transformation}. This same gauge variation is also the source of the \emph{consistent anomaly}, of the Weyl fermion,
\be{rltopcurr}
\partial_\mu J_{+{\rm con}}^\mu =  \frac e {96\pi^2} \emn F_{\mu\nu} F_{\sigma\lambda} \, .
\ee

The current Eq.~\eqref{constcurr} is conserved but not gauge invariant and, as already mentioned, we need the $AF$ term to get the  \emph{covariant} current Eq.~\eqref{aatheta_covariant}, which is however not conserved (the terminology is based on the non-Abelian case where the currents are gauge covariant rather than gauge invariant). Although the $AF$ piece in Eq.~\eqref{aatheta_covariant} cannot be be obtained from a 4D action, it does follow from the  $AAA$ triangle diagram if all vertices are treated symmetrically as is natural in the context of an effective action. Alternatively we can just  add this ``Bardeen-Zumino" $AF$ term by hand in order to get a covariant current. We note that when adding the two chiral components, the $AF$ term will cancel so total charge conservation is maintained, and we will discuss the anomalies in the individual chiral components later. 

It is well understood that the $AF$ part of the current Eq.~\eqref{aatheta_covariant}  \emph{can} be obtained as a functional derivative if we add a  {\it five\/} dimensional CS action
 \be{EQ:5d_action}
 S[\theta_\pm,A]&=&  \mp  \frac{1}{192\pi^2} \int_{M_4=\partial M_5}  d^4x\, \theta_\pm \, \epsilon^{\mu\nu\sigma\tau} F_{\mu\nu}F_{\sigma\tau} \nonumber
 \\
  &\mp&  \frac{1}{96\pi^2}  \int_{M_5} d^5x\,  \epsilon^{\mu\nu\rho \sigma\tau}A_\mu F_{\nu\rho}F_{\sigma\tau} \nonumber
  \\
  &+& S_{\mathrm {non\,top}}   [\partial_\mu \theta_\pm - 2e A_\mu] \, ,
 \ee
 where we also included a gauge invariant  non-topological term $S_{\mathrm {non\,top}}   [\partial_\mu \theta_\pm - 2e A_\mu] $.
This  action \emph{is}  invariant under the gauge transformation Eq.~\eqref{EQ:gauge_transformation}, and  the functional derivative of  Eq.~\eqref{EQ:5d_action} with respect to $A_\mu$ gives a current  on the 4D space-time boundary $M_4$ of  $M_5$, 
 \be{EQ:4d-number-current}
\hspace{-0.5cm}
J_{+}^\mu 
= - 2e \frac {\delta  S_{\mathrm {non\,top}} } {\delta A_\nu}+\frac{e}{48\pi^2} \epsilon^{\mu\nu\sigma\tau}( \partial_\mu \theta +2eA_\nu)F_{\sigma\tau} \, .
\ee
The $2eA_\nu F_{\sigma\tau}$ part of the  current Eq.~\eqref{EQ:4d-number-current} comes from the total derivative second  term in 
\be{EQ:variation_current}
&&\delta_{A_\mu}\left(-\frac 1{96 \pi^2}\int_{M_5} d^5x \epsilon^{\mu\nu\rho \sigma\tau}A_\mu F_{\nu\rho}F_{\sigma\tau}\right)\nonumber
\\
&=& \frac{3}{96\pi^2} \int_{M_5} d^5x \epsilon^{\mu\nu\rho \sigma\tau}(\delta A_\mu) F_{\nu\rho}F_{\sigma\tau} \nonumber
\\
&+&
\frac{4}{96 \pi^2} \int_{M_5} d^5x\,\partial_5\left[ \epsilon^{5\mu \nu \sigma\tau}(\delta A_\mu) A_\nu F_{\sigma\tau}\right]\, ,
\ee
and a direct calculation shows that the action  Eq.~\eqref{EQ:5d_action} is indeed gauge invariant. Since we have now shown that the topological part of the electromagnetic  current  originates from a topological action determined by the 4D anomaly, it follows that it is an exact result that does not rely upon one-loop perturbation theory as in the first derivation. 

We conclude this section with a  comment on the chiral anomaly. Without a mass (or pairing) term, the \emph{covariant} chiral anomaly is,
\be{EQ:anom}
\partial_\mu J_{+ \mathrm{cov}}^\mu= \frac{e^2}{32\pi^2} \epsilon^{\mu\nu\sigma\tau} F_{\mu\nu}F_{\sigma\tau}
\ee
and we do not expect this to change because the addition of a mass. This is also consistent with  the first term in Eq.~\eqref{EQ:variation_current} which describes the inflow  of charge from the 5D bulk and guarantees charge conservation for the separate chiralities \cite{callan1985anomalies}. The crucial factor of 3 appear since all three terms in $AFF$ on the left side contribute to the variation. Using a slight generalization of the argument in \oncite{sl} we now show that the current  Eq.~\eqref{EQ:4d-number-current} indeed has the same anomaly. For this we  first calculate the divergence of the non-topological part of the current using the $\theta$ equation of motion:
\be{eomth}
 \frac {\delta} {\delta \theta} \left(  S_{\mathrm {non\,top}}  + S_{\theta; \pm} \right) =0
 \ee
 which, using the functional form  $S_{\mathrm {non\,top}}  [\partial_\mu \theta - 2e A_\mu]$ gives 
\be{eomth2}
&-& \partial_\mu  \frac {\delta S_{\mathrm {non\,top}} } {\delta \partial_\mu \theta}  
= \frac 1 {2e} \partial_\mu  \frac {\delta S_{\mathrm {non\,top}}}  {\delta A_\mu } \nonumber
\\
&=& \partial_\mu j^\mu_{\mathrm {non\,top}} =  \frac{e^2}{96\pi^2} \epsilon^{\mu\nu\sigma\tau} F_{\mu\nu}F_{\sigma\tau}
\ee
Combining this with the divergence of the expression  Eq.~\eqref{EQ:4d-number-current}, proves that $J_{+}^\mu $ has the expected anomaly Eq.~\eqref{EQ:anom}.

%%%%%%%%%%%%%%%%%%%%%%%%%%%%%%%
\subsection{A low-energy action for the TSC  }   \label{sub:efftheory}

Our physical superconductor has both left and right-handed Weyl fermions. We already mentioned that this means that the problematic anomalies  cancel between them, ensuring that  there is no longer any need for  a five dimensional action to preserve gauge invariance.  The resulting {\it total\/} topological current is 
\be{emtopcurr}
j_{t}^\mu   =   \frac {e} {48\pi^2} \emn\left( \partial_\nu \theta_+ - \partial_\nu\theta_-\right)F_{\sigma\lambda} \, ,
\ee
which is the same as derived in \oncitec{sl} but here we obtained it directly form the original superconducting  Hamiltonian using a minimum of theoretical machinery. 

Turning to the full TSC action, a simple choice for the non-topological part is a  relativistic Abelian Higgs $\sigma$-model, but one can alternatively use a more realistic non-relativistic description (see, for instance \oncite{HANSSON2004497} for the relation between these two descriptions). With this the final low energy action for the TSC becomes,
\be{lowtheory}
S_{\text{eff}} &=&  \int d^4x \, \left[ - \frac 1 4 F_{\mu\nu}F^{\mu\nu}  - \frac 1 {192\pi^2}  \left(\theta_+ - \theta_-\right) \emn F_{\mu\nu} F_{\sigma\lambda} \right. \nonumber
\\
&+& \left. \frac {\rho_+} 2 \left(\partial_\mu \theta_+ - 2e A_\mu\right)^2 + \frac {\rho_-} 2 \left(\partial_\mu \theta_- - 2e A_\mu\right)^2  \right]  ,
\ee
where we also added the Maxwell term. As in the TI case,  the $\theta$-angles for the two Fermi surfaces must differ by $\pi$ in order to have  TRI state. At the level of the effective theory this has to be postulated, while in a non-interacting lattice model it can be inferred from the band topology. As in \oncite{qwz} we could also add a Josephson term $\sim \cos (\theta_+ - \theta_-)$  to confine vortices with non-zero chiral winding number. 

%%%%%%%%%%%%%%%%%%%%%%%%%%%%%%%%%%%%%%%%%%%%%%%%
%%%%%%%%%%%%%%%%%%%%%%%%%%%%%%%%%%%%%%%%%%%%%%%%
\section{Is there a quantum Hall effect at the surface of a  TSC?} \label{sec:cstheory}

%%%%%%%%%%%%%%%%%%%%%%%%%%%%%%%
\subsection{A naive surface theory and its difficulties}  \label{sub:naive}

We already mentioned that any measurable  topological effect in a TSC must take place at boundaries or defects since the bulk is in a Mei\ss ner phase. The most naive way to arrive at a boundary theory is to describe the interface between a TSC and a trivial SC, or an insulator  as  a step in $\theta = \theta_+ - \theta_-$. Taking $\theta = \pi \Theta(-z)$, where $\Theta$ is the step function, puts  the interface in the $xy$-plane at $z=0$, and the TSC in the $z\le 0$ half space. Putting this in Eq.~\eqref{lowtheory} and integrating by parts gives the CS term
\be{csaction}
S_{\text{CS}} = \frac {e^2} {48  \pi} \int dt d^2 r\, \epsilon^{\mu\nu\sigma} A_\mu \partial_\nu A _{\sigma}
\ee
implying  that the interface has a quantized Hall conductivity, 

\be{EQ:Hall-conductivity}
\sigma_{xy} =\frac 1{12} \frac{e^2}{h},
\ee
similar to, but only $1/6$ as large, as that  at the  surface of  a TI \cite{PhysRevLett.95.226801,RevModPhys.82.3045,RevModPhys.83.1057}.

This result is indeed puzzling, since a direct calculation based on the original Hamiltonian Eq.~\eqref{ham} restricted to a slab yields a single \emph{neutral}
Majorana surface mode, while in TI case there is a charged Dirac surface mode consistent with having a CS surface action with a half integer level number. 

Even disregarding the above  there is a tentative problem in principle with having  a quantum Hall response at  the surface in the presence of an  Abrikosov vortex carrying flux $\Phi=2\pi \hbar/2e$. A simple consideration with $\sigma_{xy}= e^2/12 h$
together with the  Streda relation $\rho = \sigma_{xy}B $, 
 shows that there is an  accumulation of electric charge $\Delta q = e/24 $ in the vicinity of the point where the vortex exits the superconductor. The corresponding result for a TI interface is $\Delta q = e/4$ as derived in \oncite{PhysRevLett.117.167002}.  Since in a SC we should identify states differing by a charge $2e$, it looks like there is a $Z_{48}$ symmetry rather than the expected $Z_{16}$ \cite{PhysRevB.93.075135}. Note that if the coefficient in Eq.~\eqref{lowtheory} had been three times larger, as claimed in \oncite{qwz}, it would, as we believe by accident,  have been consistent with the $Z_{16}$ symmetry.

A third problem is directly related to the topological current Eq.~\eqref{emtopcurr} in the presence of a vortex line and a constant electric field parallell to it. For such a configuration there is a steady radial current directed towards the vortex, as described in Section \ref{sec:res/sum}. However, just as there is no charged 2D mode at the surface, the 1D mode in the vortex core is also a neutral Majorana mode. Thus it appears that charge will just accumulate on the vortex core. 

The first two of  these puzzles would clearly be resolved if there in fact is no CS term Eq.~\eqref{csaction} in the surface action, and, as we show in the next subsection, this turns out to be the case at least for one concrete model for the interface. The problem of charge accumulation on a vortex core remains, and we shall address it in Section \ref{sec:res/sum}. 

%%%%%%%%%%%%%%%%%%%%%%%%%%%%%%%
\subsection{A better theory}  \label{sub:better}

Clearly the considerations in the previous section were naive in the sense that the current in Eq.~\eqref{emtopcurr} and the corresponding action, Eq.~\eqref{lowtheory}, were derived assuming a space without any defects or boundaries. This strategy was successful in the case of TIs, but is by no mean obviously justified. In \oncite{mb} Mulligan and Burnell set out to give such a justification in the case of a TI. More specifically, they asked whether CS term extracted from the $\theta$-term in the effective action should be added to what can be calculated directly from the surface modes, or whether these were just two alternative ways to derive the same result. The result of their calculation, which we will describe shortly, was that you should \emph{either} use the effective field theory result, \emph{or} the result directly calculated from the edge modes. If the same logic applies to the TSC case, we would expect that since the edge modes are neutral and does not give any Hall response, the same should be true for the effective field theory when derived correctly in the presence of defects or interfaces. 

The approach in \oncite{mb} was to directly calculate the surface action in a model with a concrete realization of the interface. We shall follow the same route, but first make some general comments. Since defects break translational invariance, the calculation of an effective action by integrating the fermions is normally very difficult. In our case we would ideally like to understand several cases:  a single Abrikosov vortex, a TSC-I or TSC-M interface and a TSC-SC interface. The case of a vortex is difficult to treat analytically, but we will comment on it in Sect. \ref{sec:res/sum}. To describe interfaces to an insulator or a metal, one must find self consistent boundary conditions for the fermions and the paring functions. Despite being possible we have not pursued this, but rather concentrated on the TSC-SC case which allows for a simple model that can be handled analytically. 

In such an interface one (but not both) of the angles $\theta_\pm$ must make a jump, or rapid transition between 0 and $\pi$. This can be modelled in two ways \cite{vv}. 
One possibility is to keep $\Delta$ real, but let it go smoothly to zero and change sign at the interface, corresponding to an abrupt $\pi$ change in the phase. The problem with
this approach is that the gap vanish at the interface, and also, since TRI is maintained, we  do not expect any Hall response.
The other way is to keep $\Delta$  in Eq.~\eqref{findirac} at some finite real value, and let the phase, say $\theta_+$, wind rapidly from $\pi$ to 0 in the interface. Such a twist breaks the TRI in the interface, and does allow for a Hall response. This realization of an interface intuitively corresponds to a regularized version of the step function used to derive Eq.~\eqref{csaction}, but we do not know how to extract an effective action in a controlled analytical way for such an interface. The model used in \oncite{mb} is similar, but not identical, to this second alternative. It amounts to letting the real part of the mass (or gap), $m$, to go smoothly to zero and change sign at the interface, while the imaginary TRI breaking part, $\sim i \sigma\gamma^5 $, is kept constant. This means that TRI in bulk is only recovered in the limit of $\sigma$ going to zero. This is the model we shall use also for the TSC.

It is now clear how to formulate  the problem: Calculate the surface response  generated by integrating out  fermions with an (Euclidean) action density
\be{reglagdirac}
\mathcal{L} =  \bar \Psi_\pm \left[ \gamma^\mu \partial_\mu +m(x)+i\gamma^5 \sigma \right] \Psi_\pm 
\ee
 where the fields $\Psi_\pm$ satisfy a Majorana condition so that the chirally twisted mass is a {\it Majorana mass\/}. 

Comparing with Eq.~\eqref{lagdirac}, the gap function is $\Delta = \sqrt{m^2(x) + \sigma^2}$ and the twist angle $\theta = \sgn(\sigma) \tan^{-1} \frac \sigma {m(x)}$.
If we take  the interface to be a  plane  and use the the profile
\be{EQ:profile}
m(x) = m_0 \tanh m_0 x \, ,
\ee
where $x$ is the coordinate perpendicular to the plane, and  $m_0\geq 0$, the total winding in the angle $\theta$ as we pass though the interface is 
\be{EQ:winding_angle}
\Delta \theta = \sgn(\sigma) 2 \tan^{-1} \left(\frac {m_0} \sigma\right) \, .
\ee
If the formula Eq.~\eqref{emtopcurr} derived from the putative effective theory is correct, a constant electric field $E_x$ at the interface, should give rise to a  Hall current, 
\be{EQ:Interface-current}
I^y_{\rm total}= - \frac{\sgn(\sigma)}{12\pi^2}  \tan^{-1} \left(\frac {m_0} \sigma\right) E_x \, ,
\ee 
corresponding to an effective Hall conductivity,
\be{hallcond2}
\sigma_{yx} =  \frac{\sgn(\sigma)}{6\pi}  \tan^{-1} \left(\frac {m_0} \sigma\right) \, .
\ee
which of course agrees with  Eq.~\eqref{EQ:Hall-conductivity}   in the limit $\sigma\rightarrow 0$. 

\begin{figure}
\includegraphics[width=\linewidth]{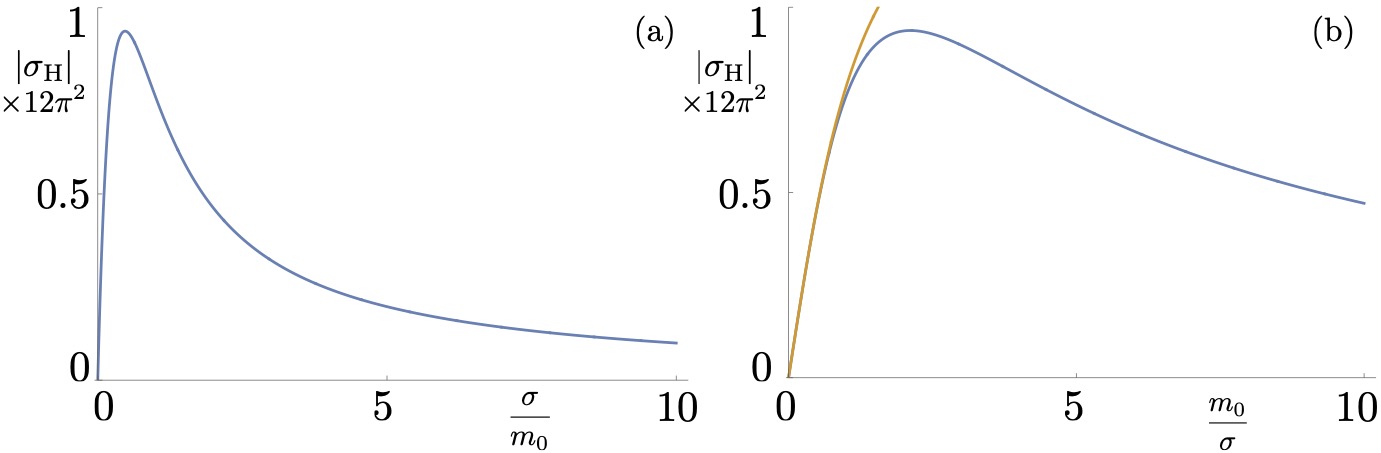}
\caption{The Hall coefficient $\sigma_H$  as (a) a function of $\sigma/m_0$ for small $\sigma$ and  and (b) as a function of  $m_0/\sigma$ for large $\sigma$.  In (b) we also show the effective theory result $I_{\rm total}= \frac{\sgn(\sigma)}{12\pi^2}   \tan^{-1} \left(\frac {m_0} \sigma\right)$  in yellow. }
%\label{fig:surface_current_TSC}
 \label{fig:hallcoef}
\end{figure}

In the next section we will show how to compute the Hall current starting from from the Majorana mass action, Eq.~\eqref{lagdirac}. We have not been able to calculate the current density, so we do not know the exact $z$-profile, but very likely it is exponentially localized close to the surface for any finite value of $\sigma$. 
Figure \ref{fig:hallcoef} (b) shows the effective Hall conductivity $\sigma_{yx} = I^y/E_x$ as calculated in the next section as a function of  $\sigma/m_0$, and we see that the large values for large values the result agrees with the effective theory prediction shown in the same graph. In Figure \ref{fig:hallcoef} (a), we show the opposite limit of small $\sigma/m_0$, and we see that, as opposed to the TI case, the Hall conductivity vanish as $\sigma\rightarrow 0$. The coefficient in front of the linear term at $\sigma = 0$ is  $-6.06247$ in units of $1/12\pi^2$.

It is at first sight not surprising that the formula  extracted from  the TSC effective action  for the Hall current is not valid when the phase angle winds rapidly. The entire effective action formalism is based on an adiabatic  approximation in which  the mass of the integrated-out Fermi field is supposed much larger than the space or time rate of change of the background fields.
On the other hand, it was shown in \oncite{mb} that in the TI case, the effective field theory result is in fact independent of the speed of the winding. One might wonder why, but in
this latter case there is indeed a good reason.  Imagine a magnetic field perpendicular to the direction of the interface so that Landau levels are formed. In the massless case, this is the setting for the  ``infinite hotel" derivation of the axial anomaly \cite{nielsen1983}, which is entirely due to the lowest Landau level (LLL). In the massive case this level maps onto a 1D fermion with a chiral mass, so if we start from an untwisted interface and slowly twist the $\theta$ field we can appeal to the argument of Goldstone and Wilczek\cite{PhysRevLett.47.986} that the total charge accumulated in the vicinity of the twist is $- \Delta \theta/2\pi$ even though the large mass-gap formula $\rho= - \partial_x \theta/2\pi$ is no longer valid. This is because the  corrections to this formula are total derivatives. In other words the total charge is a topologically protected quantity. Thus, since a filled LL gives the Hall coefficient $1/2\pi$, in the TI case we expect a Hall coefficient $\Delta\theta/2\pi$ corresponding to a CS level 1/2 for a $\pi$ twist. 

There is no corresponding argument in the TSC case. In the presence of a mass the axial LLs behave very differently from the usual vector ones, and in particular there is no simple mapping of the lowest degenerate energy level to a 1D fermion.  We discuss this case in some more detail in Appendix \ref{app:axial}, but it seems clear that in the TSC case the ``topological" $\theta$-term is \emph{not} so topologically protected.

%%%%%%%%%%%%%%%%%%%%%%%%%%%%%%%
\subsection{The Chern-Simons action in a TSC-SC interface -- a perturbative analysis.} \label{sub:surfaceact}

As shown in \oncite{mb}, the profile Eq.~\eqref{EQ:profile} allows for a closed analytic form for the fermion propagator, and following this reference we shall decompose it one term localized to the interface, \ie $x_4 = 0$, and another describing the scattering states (note that our notation differ from that in \oncite{mb} since we work in Euclidean space). We use a mixed representation $(x;k^i)$ where $x=x_4$ and $k^i$ is the three momentum where $i =1,2,3$.   With this we have
\be{fullprop}
S(k^i ; x,x') = S_{\text{l}} (k^i ; x,x') + S_{\text{f}} (k^i ; x,x')
\ee
where the localized part is,
\begin{widetext}
\be{locprop}
S_{\text{l}} (k^i ; x,x') = \int \frac{ds} {2\pi} \,  \frac 1 {k^2 + \sigma^2}  \frac 1 {s^2 +k^2 + M^2} P_+ (i k_i \g^i + i \sigma \g^5) \mu(s; x,x') e^{is(x-x')}
\ee
and the scattering, or free, part is
\be{freeprop}
S_{\text{f}} (k^i ; x,x') =  - \int \frac{ds} {2\pi} \,  \frac 1 {s^2 +k^2 + M^2} ( i s\gamma^4 + i k_i \g^i + i \sigma \gamma^5 - \tilde M(x,x') ) e^{is(x-x')}
\ee
where $M^2 = m_0^2 + \sigma^2$,  $ \tilde M(x,x') = P_+ m(x) + P_- m(x') $, with the projectors $P_\pm = \half (1 \pm \gamma^4)$, and 
$$
\mu (s; x,x') = is ( m(x) - m(x' ))  + m(x) m(x') - m_0^2 \, .
$$
The expressions Eqs.~\eqref{locprop} and \eqref{freeprop} were given in \oncite{mb}, and in Appendix \ref{app:prop} we provide a somewhat simpler derivation. 

With this we can extract the effective action to ${\mathcal O}(A^2)$, \ie terms $\sim F\tilde F$ or $AdA$ from the polarization diagram in Figure \ref{fig:diag}.
We will in the following only care about terms that contribute to the CS action $AdA$ which originate from terms with the $\gamma$-trace structure $\mathrm{Tr} \left( \g^i\g^5 \g^4 \g^k q_k \g^j\right)$. The relevant trace coming from the $S_{\text{l}} S_{\text{l}}$ contribution to $\Pi^{ij}$ is,
\be{tzerotrace} 
&\mathrm{Tr} \left\{ \g^i\g^5 P_+  (i k_l \g^l + i \sigma \g^5)\g^j\g^5 P_+ \left[i (k -q)_l \g^l + i \sigma \g^5\right] \right\}  \\ 
&= \mathrm{Tr} \left\{ \g^i\g^5 P_+ P_- (i k_l \g^l + i \sigma \g^5)\g^j\g^5 \left[i (k -q)_l \g^l + i \sigma \g^5\right] \right\} = 0 \, , \nonumber
\ee
\end{widetext}
where the first identity follows from the identities $\g^i P_\pm = P_\mp \g^i$ and $\g^5 P_\pm = \pm P_\mp \g^5$ and the second from $P_+P_- = 0$. Note that for a vector coupling this trace would not be zero, and it in fact gives the CS term in the case of a TI as shown in \oncite{mb}. 

In Appendix \ref{app:poltens} we evaluate the remaining contributions, $S_{\text{l}} S_{\text{f}}$, $S_{\text{f}} S_{\text{l}}$ and $S_{\text{f}} S_{\text{f}}$, and show that, although non-zero, they also do not give any CS term in the surface action in the TRI limit $\sigma \rightarrow 0$.

Having shown that there is no quantum Hall response in an interface where a constant $\sigma$ is taken to zero, can we be assured that the same holds also
for  other types of interfaces? We could \eg take $\sigma\ne 0$ only close to the interface, thus keeping the gap, or just put $\sigma = 0$ and break TRI in some other ways, \eg by some magnetic impurities. We cannot logically exclude the  possibility that a CS term is generated in these cases, but a numerical study of the analogue (1+1)D case in \oncite{Spanslatt2015},
indicates that, at least qualitatively, the physics of $\pi$-junctions do not depend on such microscopic details.

%%%%%%%%%%%%%%%%%%%%%%%%%%%%%%%%%%%%%%%%%%%%%%%%
%%%%%%%%%%%%%%%%%%%%%%%%%%%%%%%%%%%%%%%%%%%%%%%%

 \section{Physical consequences and summary of results} \label{sec:res/sum}
 
The result of the  last section resolves two of the problems posed in Section \ref{sub:naive}.  Since there is no CS term in the surface action, there is no contradiction with the localized surface mode being neutral. There will also not be any charge accumulation associated to magnetic flux lines, and thus no problem with the expected $Z_{16}$ symmetry of the TSC. Note, however, that this is true only in the limit $\sigma \rightarrow 0$; both effects remain for finite symmetry breaking. 
For the following discussion  we rewrite the action Eq.~\eqref{lowtheory} as
\be{lowtheory2}
S_M &=&  \int d^4x \, \left[  \frac \rho 2 \left(\partial_\mu \vartheta + 2 A_\mu \right)^2 + \frac \rho 8 \left(\partial_\mu\theta\right)^2 + J \cos\theta \right. \nonumber
\\
&+& \left.\frac \theta {96\pi^2} \epsilon^{\mu\nu\sigma\lambda} F_{\mu\nu} F_{\sigma\lambda} - \frac 1 4 F_{\mu\nu}F^{\mu\nu} \right] 
\ee
where we,  following \oncitec{qwz} also added a Josephson term which provides a gap for the $\theta$ mode that is not gapped by the Higgs mechanism, and where
$$ \vartheta = \half \left(\theta_+ + \theta_-\right) \ \ \ \ \mathrm{and} \ \ \ \ \theta=\theta_+ - \theta_-$$
and where we put $\rho_R = \rho_L = \rho/2$. The corresponding full electromagnetic current is
\be{fullcurrent}
J^\mu = \rho \left(\partial^\mu \vartheta + 2 A^\mu\right) + \frac e {48\pi^2} \epsilon^{\mu\nu\sigma\lambda }\partial_\nu \theta F_{\sigma\lambda} \, .
\ee

We next calculate the charge inflow towards a straight thin (\emph{i.e.}, the limit of the London length $\lambda_L \rightarrow 0$) Abrikosov vortex line, in the presence of a parallell constant electric field so,
\be{fields}
\vec B = B \, \delta^2 (\mathbf{r} ) \hat z ;\quad \vec E = E \hat z
\ee
with $E$ and $B$ constant. In the $A_0  = 0$ gauge, and using radial coordinates, $\mathbf {r} =  (r, \alpha)$ we take the potentials, 
\be{potentials}
A_r = 0;\quad A_\alpha = \frac {\phi_0} {2\pi} \hat \alpha; \quad A_z = - E_z t \hat z \, .
\ee
By the equation of motion for $\vartheta$, the term $\sim \rho$ in \eqref{fullcurrent} vanishes, while the second gives a gauge invariant radial current,
\be{radialj}
J^r = \frac e {48\pi^2} \frac 1 r  \partial_\alpha \theta \, 2 E_z
\ee
and thus  the puzzling charge accumulation (per unit length) on the vortex core mentioned in the introduction is, 
\be{chacc}
\dot Q = \oint d\alpha\, r J_r  = \frac e {12\pi} E_z \, .
\ee
Since the current is conserved, such an accumulation should take place  if there is no charge transported along the vortex.  For completeness, we show how this formally comes about. To do so, recall that the vortex we consider is in one of the two phase fields, but not in both. For concreteness we take a unit vortex in $\theta_+$ and no vortex in $\theta_-$ so $\vartheta = \half \theta_+$ and $\theta = \theta_+$. Assuming that the total current in the $z$-direction vanishes, which is the case if there is no charged mode in the vortex, the equation of motion for $\vartheta$  gives $\partial_z \theta = 2 \partial_z \vartheta = -4 A_z$. Substituting this in Eq.~\eqref{fullcurrent} gives for the $0^{th}$ component,
\be{topchden}
J^0 &=& - \frac {e^2} {96\pi^2} \partial_z \theta \, 2  B \, \delta^2 (\mathbf{r} ) = -  \frac {8 e^2} {96\pi^2}  B \, \delta^2 (\mathbf{r} )  A_z \nonumber
\\
&=&   \frac { e^2} {12\pi^2}  B \, \delta^2 (\mathbf{r} )  E_z \, t \, .
\ee
Integrating over $\mathbf {r}$ and taking the time derivative we recover Eq.~\eqref{chacc}. \\

In the small $\sigma$ limit, the absence of a charged mode in the vortex is closely related to the absence of a Hall effect in a TSC-SC interface. To see this consider a geometry where the vortex line is fully inside the cylinder of normal SC embedded in the TSC. If there was a CS response in the interface, this could ensure current conservation without any charge accumulation in the boundary layer.\\

As already mentioned, for finite $\sigma$ we both have a (non-quantized) Hall effect in interfaces, and an unphysical charge accumulation on Abrikosov vortices in the $\theta$ variable. At least the latter is clearly unphysical. There are two possible way to resolve this conundrum. The most obvious one is to note that the assumption of having a constant electric field in a superconductor is clearly unphysical, since it will be screened by the Cooper pair super current. Similarly, as soon as a Maxwell term is present,  a buildup of  charge is not possible since it will again be screened by the Cooper pairs. In both cases the screening becomes local in the limit of large electric charge.

We might, however, consider the opposite limit, $e \rightarrow 0$, while keeping $eE_z$ fixed, where screening effects are neglected. In this case the non-intuitive charge build up seems to take place. An appealing resolution would be that there is some principle that enforces $\theta=0$, in which case non of the strange effects would be present even in the absence of screening. Ideally there would be some robust geometrical or topological reason for this condition, but we have not been able to find any. There is, however, an energetic reason for setting $\theta=0$. Without the Josephson term in Eq.~\eqref{lowtheory2}, the  $\theta$-phase degree of freedom would be massless, and since the originally model Eq.~\eqref{ham} is fully gapped, there is a strong reason to believe that $J\ne 0$. Since $\theta$ by construction is periodic with period $2\pi$, states with $\theta = 0$ and $\theta = \pi$ cannot simultaneously be ground states of the model. This seems to rule out a $\pi$-junction in $\theta$, since the energy cost is not proportional to the area of the interface but rather to the volume of the bulk.  It is also clear that energetics will severely restrict vortex solutions in $\theta$ since there is no Higgs mechanism, a single vortex has the usual logarithmic divergence in the energy per unit length, and this will remain true in the presence of  a  small Josephson coupling. There will not be any net charge accumulation around a vortex-antivortex pair, but it is also not clear if there is a stable ground state in such a configuration without invoking screening. 

An alternative interpretation of our results is that when we have an apparently anomaly-free system of two opposite-chirality Weyl fermions,  each gapped with its own complex Majorana-mass order parameter, then there is a novel source of anomaly such that the  electromagnetic current is conserved only when the phases $\theta_\pm$ of the two mass terms  differ by a constant.  Such a  constraint,  independent of any energetic arguments, would  make the problematic transverse current vanish identically. At the moment we do not have an explanation as to how such an anomaly would arise

\vskip 2mm
\noi{\bf Acknowledgement}
T.H. Hansson and M. Stone thank Mike Mulligan for  helpful discussions about \oncitep{mb}  T.H. Hansson  thanks Yoran Tournois for collaboration at an early stage of this project, 
and he  also thanks the Yukawa Institute for Theoretical Physics for hosting him during the period when this work was initiated.
 M. St\aa lhammar thanks Lukas R\o dland for fruitful discussions. M. St\aa lhammar and T.H. Hansson are supported by the Swedish Research Council (VR) and  M. St\aa lhammar also by the Wallenberg Academy Fellows program of the Knut and Alice Wallenberg Foundation. M. Sato is supported by JSPS KAKENHI (Grant No. JP20H00131), JST CREST project (No. JPMJCR19T2), and JSPS Core-to-Core Program (No. JPJSCCA2017002).

%%%%%%%%%%%%%%%%%%%%%%%%%%%%%%%%%%%%%%%%%%%%%%%%
%%%%%%%%%%%%%%%%%%%%%%%%%%%%%%%%%%%%%%%%%%%%%%%%

\begin{widetext}
 \appendix
\section{Derivation of the fermion propagator}  \label{app:prop}
\def\sech{{\rm sech\,}}

In this Appendix we derive the expressions Eqs. \eqref{locprop} and \eqref{freeprop} for the localized and free parts of the fermion propagator  given in the main text.
This amounts to compute the Green's function of the Euclidean signature Dirac operator 
\be{defd}
D= \gamma^\mu \partial_\mu +m\left(x\right)+i\gamma^5 \sigma
\ee
where $x=x_4$.

Just as the free space Dirac Green's function is calculated as $(\slashed  p -  m) \times (p^2 + m^2)^{-1}$, our strategy is to first compute the Green's function of $DD^\dagger$ and then use 
\be{trick}
D^\dagger  [DD^\dagger]^{-1}= D^\dagger [D^\dagger]^{-1} D^{-1}= D^{-1} \, .
\ee 
\be{gg}
DD^\dagger&=& \left[\gamma^\mu \partial_\mu +m(x)+i\gamma^5 \sigma\right]\left[-\gamma^\mu \partial_\mu +m(x)-i\gamma^5 \sigma\right]\nonumber \\
&=& - \partial^2_{x}  +\gamma_4 \partial_{x}m(x)  + m^2(x) +\sigma^2 - [\nabla^2]_\perp \\
&\to& - \partial^2_{x}  +\gamma_4 \partial_{x}m(x)  + m^2(x)+ \kappa^2 \nonumber
\ee
where $\kappa^2= k_\perp^2+\sigma^2$. We now take the profile  $m(x)=m_0\tanh m_0 x_4$ to get 
\be{defm}
m^2(x_4)= 1-{\rm sech}^2 x_4, \quad \partial_x m(x_4)={\rm sech}^2 x_4 \, , 
\ee
and thus, 
\be{dddag}
DD^\dagger  = - \partial^2_{x_4} - (1-\gamma_4) {\rm sech}^2 x_4 +1+\kappa^2 \, .
\ee
Here, and in the following, we shall put $m_0 = 1$ and restore the dimensions only at the end. 
The calculation simplifies by taking $\gamma_4$ to be diagonal, and we shall use,
\be{gamma4}
\gamma_4 = \begin{pmatrix} {\mathbb I}_2 & 0 \\  0 &-{\mathbb I}_2 \end{pmatrix} 
\ee
giving 
\be{xxx}
DD^\dagger= \begin{pmatrix}  - \partial^2_{x_4}  +1+\kappa^2 &0 \\  0 &  - \partial^2_{x_4} - 2 {\rm sech}^2 x_4 +1+\kappa^2 \end{pmatrix}
\ee
which contains a pair of the P\"oschl-Teller equations.
The corresponding eigenfunctions of $DD^\dagger$ are 
\be{scsol}
        \psi_{k,{\rm sc}}({\bf x}) = \begin{pmatrix}1 \\  \frac {(-ik_4+\tanh x_4)} {\sqrt{1+k_4^2}} \end{pmatrix} e^{i{\bf k} \cdot {\bf x}}, \quad \lambda = k_4^2+ 1 + k^2_\perp+\sigma^2,
\ee
and 
\be{boundsol}
\psi_{\rm b}({\bf x}) = \begin{pmatrix} 0 \\ \frac 1 {\sqrt 2} {\sech  x_4} \end{pmatrix} e^{i{\bf k}_\perp \cdot {\bf x}_\perp}, \quad \lambda= k^2_\perp+\sigma^2.
\ee
which satisfy the completeness relation 
\be{complete}
\delta^4({\bf x}-{\bf x}')= \int \frac{d^3 k_\perp}{(2\pi)^3} \left[ \psi_{\rm b}({\bf x}) \psi_{\rm b}^\dagger({\bf x}') + \int  \frac{dk}{2\pi}  \psi_{k,{\rm sc}}({\bf x}) \psi^\dagger_{k,{\rm sc}}({\bf x}')\right] \, .
%e^{i{\bf k}_\perp\cdot ({\bf x}-{\bf x}')_\perp}
\ee
The desired $[DD^\dagger]^{-1}$  Green function is then nothing but the completeness relation divided by 
the eigenvalues.     
To avoid clutter we will not write explicitly the factors of  $[d^3k_\perp/(2\pi)^3]e^{i{\bf k}_\perp\cdot ({\bf x}-{\bf x}')_\perp} $, and with this understood the bound state contribution to  $[DD^\dagger]^{-1}$ is
$$
{\textstyle \frac 12}  (1-\gamma_4) \frac { \psi_{\rm bound}(x)\psi_{\rm bound}(x')}{\lambda_{\rm bound}}={\textstyle \frac 12}  (1-\gamma_4) \frac{{\textstyle \frac 12} \sech x \,\sech x'}{k^2_\perp+\sigma^2} \, ,
$$
and the scattering state contribution,
$$
{\textstyle \frac 12} (1-\gamma_4) \int \frac {d k}{2\pi } \frac{(-ik +\tanh x)(ik +\tanh x') e^{ik(x-x')}}{(1+k^2)(1+k^2+k_\perp^2 +\sigma^2)}+{\textstyle \frac 12} (1+\gamma_4) \frac{e^{ik(x-x')}}{1+k^2+k_\perp^2 +\sigma^2} \, .
$$ 
To get the fermionic $D^{-1}$ Green function we  then apply 
$
D^\dagger =-\gamma^\mu \partial_\mu +m(x)-i\gamma^5 \sigma
$
to these contributions.  

The bound state part is easily shown to be,
\be{boundprop}
D_{\rm bs}(x,x')=-{\textstyle \frac 12} (1-\gamma_4) \frac{\gamma^a k_a + i\gamma_5 \sigma}{k_\perp^2+\sigma^2}  {\textstyle \frac 12} \sech x \,\sech x' \, .
\ee
while it requires a bit more work to get the scattering part. This involves using,
\be{relation1}
\left(\partial_x +\tanh x\right)\left(-ik +\tanh x\right)e^{ikx}&=& \left(1+k^2\right)e^{ikx} \nonumber \\
\left(\partial_x +\tanh x\right) \sech x&=&0 \, ,
\ee
and following \oncite{mb} we also introduce the function  $\mu(x,x)$ {\it via\/}
\be{relation2}
\frac{(-ik +\tanh x)(ik +\tanh x')}{k^2+1}&=& 1+ \frac  1{k^2+1} \left[  ik\left(\tanh x-\tanh x'\right)+\tanh x \,\tanh x-1  \right]  \\
&\stackrel{\rm def}{=}&1+ \frac{1}{1+k^2}\, \mu(x,x') \nonumber \, .
\ee
Doing the algebra we get for the scattering part
$$
D_{\rm sc}(x,x')=- \int \frac {d k}{2\pi } \frac{\gamma^\mu k_\mu + M(x,x') +i\gamma_5 \sigma}{k^2+1+ k_\perp^2 +\sigma^2} e^{ik(x-x')},
$$
where 
\be{defbigm}
M(x,x')= -{\textstyle \frac 12} \left(1-\gamma_4\right)\tanh x'  -{\textstyle \frac 12} \left(1+\gamma_4\right)\tanh x+ {\textstyle \frac 12} \left(1-\gamma_4\right) \frac{\gamma^a k_a+i\gamma_5 \sigma}{k^2+1}\mu(x,x') \, .
\ee
%and where we recall that $\mu(x,x') = ik (\tanh x-\tanh x') +\tanh x\, \tanh x' -1.$
Here we note that the term $\sim \mu(x,x') $ is localized to the surface, and we shall combine that with the bound state propagator Eq. \eqref{boundprop}.
To do this, we first make the partial fraction decomposition, 
\be{partdec}
\frac{1}{\left(k^2+1\right)\left(k^2+1+ k_\perp^2 +\sigma^2\right) }= \frac{1}{k_\perp^2+\sigma^2}\left( \frac 1{ k^2+1}-\frac 1{k^2+1+ k_\perp^2 +\sigma^2}\right)
\ee 
and then use the relation
\be{relation4}
\int \frac{dk}{2\pi} e^{ik(x-x')}  \frac{\mu(x,x')}{1+k^2}= -{\textstyle\frac 12}\sech x\,
\sech x' \, ,
\ee
that is given on p. 13i  in \oncite{mikebook}. We then see that the contribution from the first term in the RHS cancels $D_{\rm bs}(x,x')$, and, restoring $m_0$, the remaining part of the localized term $\sim \mu(x,x') $ gives the localized propagator $D_{\rm l}(x,x')$ in  Eq. \eqref{locprop}, while the remaining part of $D_{\rm sc}(x,x')$ gives $D_{\rm f}(x,x')$ in Eq. \eqref{freeprop}.  

%%%%%%%%%%%%%%%%%%%%%%%%%%%%%%%%%%%%%%%%%%%%%%%%
%%%%%%%%%%%%%%%%%%%%%%%%%%%%%%%%%%%%%%%%%%%%%%%%
\section{Vector and axial Landau levels} \label{app:axial}
In this Appendix, we prove some of the assertions made in Section \ref{sub:better}, and give plausibility arguments for some of the others. 

Our starting point is the Hamiltonians for a Dirac field coupled to a vector potential $\vec V$ or an axial potential $\vec A$, and with a general chiral mass $M = m + i\gamma^5\sigma$.
It will be convenient to use the Weyl representation for the (Minkowski space) $\gamma$-matrices, 
\be{weylgamma}
\gamma^0 = 
\begin{pmatrix}
0 & -1 \\ -1 & 0
\end{pmatrix};\quad
\gamma^i = 
\begin{pmatrix}
0 & \sigma^i \\ -\sigma^i & 0
\end{pmatrix};\quad
\gamma^5 = 
\begin{pmatrix}
1 & 0 \\ 0 & -1
\end{pmatrix} \, .
\ee
and we also introduce the Dirac matrices $\alpha^i = \gamma^0 \gamma^i$, and $\beta = \gamma^0$.  With this we can write
the massive Dirac Hamiltonians,  
\be{massham}
H_{\mathrm{V} } &= \vec\alpha\cdot (\vec p + \vec V) + \beta m + i\beta\gamma_5 \sigma \, = \begin{pmatrix} H_+ & M \mathbb {I}  \\ M^\star \mathbb { I} & H_+ \end{pmatrix} \, ,  \nonumber \\
H_{\mathrm{A} } &= \vec\alpha\cdot (\vec p + \gamma^5 \vec A) + \beta m + i\beta\gamma_5 \sigma \, = \begin{pmatrix} H_+ & M \mathbb {I}  \\ M^\star \mathbb { I} & H_- \end{pmatrix} \, .
\ee
Next,  we specialize to a constant magnetic field in the $z$-direction,  and use the symmetric gauge, $V = \frac {B_V} 2 (-y, x)$ and $A = \frac {B_A} 2 (-y, x)$, where $B_V$ and $B_A$ are the magnitudes of the vector and axial magnetic fields respectively. 

To understand the difference between vector and axial vector cases, it is useful to write the  Hamiltonian matrices explicitly in terms of $a$ and $a^\dagger$, 
\be{hvec}
H_V = \begin{pmatrix}
p_z &  \sqrt B_V a^\dagger & M & 0 \\
 \sqrt B_V a & -p_z & 0 & M \\
M^\star & 0 & -p_z & - \sqrt B_V a^\dagger \\
0 & M^\star & - \sqrt B_V a & p_z 
\end{pmatrix}
\ee
and
\be{havec}
H_A = \begin{pmatrix}
p_z &  \sqrt B_A a^\dagger & M & 0 \\
 \sqrt B_A a & -p_z & 0 & M \\
M^\star & 0 & -p_z & - \sqrt B_A a \\
0 & M^\star & - \sqrt B_A a^\dagger & p_z 
\end{pmatrix} \, .
\ee
The solution to the Landau problem is now conveniently expressed in terms of the eigenfunctions $\ket n$ of the Harmonic oscillator Hamiltonian $H = \sqrt B ( aa^\dagger + \half)$, with $[a,a^\dagger] =1$,  $H \ket n = (n+\half) \ket n$, and the plane waves $e^{ip_z z}$.

First consider the massless  case, where both Hamiltonians are block diagonal. The eigenfunctions for $H_+$ are  $\psi_+ = N_+(\ket n, \pm\frac {\sqrt {nB}}{p+E} \ket {n+1})^T$, and for $H_-$,  $\psi_l = N_- (\ket {n-1}, \frac {xx} {yy}\ket {n})^T$, and we recall that all these LLs are massively degenerate since the angular momentum lowering and rising operators commute with both Hamiltonians. We also note that in both cases there are linearly dispersing chiral modes for $n=0$. The spectral flow of these LLLs are the origin of the axial (and vector) anomalies as originally explained by Nielsen and Ninomiya \cite{{nielsen1983}}. 

Recalling that $a^\dagger \ket {n-1} = \sqrt n \ket n$ and $a \ket n = \sqrt n  \ket {n-1}$ we see by inspection  that,
\be{ansatz}
\psi_V = \begin{pmatrix} \alpha \ket n \\ \beta \ket {n-1} \\ \gamma \ket n \\ \delta \ket {n-1} \end{pmatrix} e^{i p_z z}
\ee
reduces the Schr\"odinger equation $H_V \psi_V = E \psi_V$ to a set of algebraic equations for the parameters $\alpha,\ \beta,\ \gamma, \ \delta$.  
For $n>0$ this yields two doubly degenerate solutions with $E_n(p) = \pm \sqrt{p_z^2 + nB + M^2}$. 

For $n=0$, \ie the LLL, there are only two solutions with $E_0(p) = \pm \sqrt{p_z^2 + M^2}$. In this last case, the dispersion relation is that of a massive 1D fermion, and  there is an exact mapping between the 3D and 1D problem (this is easily seen \eg  by noting that for LLL wave functions,  the second and fourth rows and columns in $H_+$ can be neglected which amounts to using a redundant parametrization of the 1D gamma matrices). 
The 1D problem was originally considered by Goldstone and Wilczek\cite{PhysRevLett.47.986}, who proved that in the presence of a chiral twist, there is a charge $\Delta Q = \sgn(\sigma) \frac 1 {2\pi} \tan^{-1} \left(\frac {m_0} \sigma\right)$ accumulated at the location of the twist. Since in  our case a unit charge amounts to a filled Landau level, we get the result for the CS coefficient claimed in the main text, if there is no contribution from the higher LLs. We do not have any rigorous proof of this assertion, but since we are clearly seeing a topological effect, it is natural to assume that only the LLL is involved, since this is the case for the axial anomaly, where the contributions from higher LLs cancel since the spectral flow of the degenerate states occur in opposite directions. 

Although the massless axial case looks very similar to the vector one, they are very different when masses are included. To illustrate this, consider the ``obvious" ansatz wave function,
\be{ansatz2}
\psi_A = \begin{pmatrix} \alpha \ket n \\ \beta \ket {n-1} \\ \gamma \ket {n-1} \\ \delta \ket {n} \end{pmatrix} e^{i p_z z}
\ee
where it is clear that the presence of a mass term will mix the LLs. It  is also  easy to convince oneself that no simple variation of this ansatz will work, and that any solution will likely involve an infinite sums over the $\ket n$ states, and this is also true for the generalized LLL. In particular, there is no obvious mapping on a 1D system. In spite of this, our observation that at large $\sigma$, \ie a slow twist of $\theta$, we do see a Hall current where the $m_0/\sigma$ behavior is consistent with a 1D interpretation just as in the vector case. We have no explanation for this, and we also note that the current related to the anomaly has the crucial factor 1/3 characteristic of the consistent anomaly, while the LL based derivation of the axial anomaly gives the covariant anomaly, where the factor 1/3 is missing.

%%%%%%%%%%%%%%%%%%%%%%%%%%%%%%%%%%%%%%%%%%%%%%%%
%%%%%%%%%%%%%%%%%%%%%%%%%%%%%%%%%%%%%%%%%%%%%%%%
 
 \section{The polarization tensor} \label{app:poltens}
 
In this Appendix, we calculate the propagator given by,
\begin{equation}
\Pi^{\mu\nu} = -\frac{1}{2} \int dxdx'\Tr\left[ \gamma^{\mu}\gamma^5S\left(x,x'\right)\gamma^{\nu}\gamma^5S\left(x',x\right)\right].
\end{equation}
For the sake of clarity, the calculation will be divided into two subsections, one dealing with the trace calculation, and one where the integrals are performed.
\subsection{Trace Calculations}
The action of the model considered can be split into,
\begin{align}
S_{\text{l}} &= \int\frac{d^3k}{\left(2\pi\right)^3}\int \frac{ds}{2\pi}\frac{1}{k^2+\sigma^2}\frac{1}{k^2+s^2+ m_0^2+\sigma^2 }P_+\left( i \gamma^ak_a + i\gamma^5 \sigma\right) \mu\left(s;x,x'\right)e^{is\left(x-x'\right)},
\\
S_{\text{f}} &= - \int\frac{d^3k}{\left(2\pi\right)^3}\int\frac{ds}{2\pi} \frac{is\g^4 + i\gamma^ak_a + i \gamma ^5\sigma - \tilde{M}\left(x,x'\right)}{k^2 +s^2+ m_0^2+\sigma^2 }e^{is\left(x-x'\right)},
\end{align}
with the identifications,
\begin{align}
\mu\left(s;x,x'\right) &:= is\left[m\left(x\right)-m\left(x'\right)\right]+m\left(x\right)m\left(x'\right)-m_0^2; \quad m\left(x\right) = m_0\tanh\left(m_0x\right)
\\
s&:=k_4; \quad P_{\pm} :=\frac{1}{2}\left(\Id + \gamma^4\right); \quad \tilde{M}\left(x,x'\right) = P_+m\left(x\right)+P_-m\left(x'\right).
\end{align}
We seek contributions to the CS term, meaning that we only consider terms of the form
\begin{equation}
\Tr\left( \g^a\g^b\g^c\g^4\g^5\right) = -4i \epsilon^{abc},
\end{equation}
with $a,b,c \in \{1,2,3\}$.
Taking this into account, we can make a first observation: there will be no contribution to the CS term from $S_{\text{f}}S_{\text{f}}$, since we cannot produce the correct combination of $\g$-matrices in any way. Let us therefore start by investigating a possible contribution from $S_{\text{l}}S_{\text{l}}$. Only the relevant terms within the trace are kept, and the equality sign should therefore be interpreted in this way.
\begin{equation}
-\Tr \left[\g^a\g^5P_+\left(\g^bk_b+\g^5\sigma\right)\g^c\g^5P_+\left(\g^dk'_d+\g^5\sigma\right)\right].
\end{equation}
By moving the left-most $P_+$ such that it reaches the right-most $P_+$, we have to use the anti-commutation relations three times, noting that $a,b,c\neq 4$. Therefore we end up with,
\begin{equation}
-\Tr\left[\g^a\g^5\left(\g^bk_b + \g^5\sigma\right) \g^c\g^5P_-P_+\left(\g^dk_d'+\g^5 \sigma\right)\right],
\end{equation}
which identically vanish since $P_-P_+=0$. This is not the case if we only have a $\g^a$-coupling, since we would have to move $P_+$ one step less than above, which results in a $P_+P_+$ instead.

Let us now move to the cross terms, and first consider $S_{\text{l}}S_{\text{f}}$. 
\begin{align}
&\quad \Tr\left[\g^a\g^5P_+\left(\g^bk_b + \g^5\sigma\right)\g^c \g^5\left(\g^dk_d'+\g^5 \sigma\right)\right] \nonumber
\\
&=\Tr\left(\g^a\g^5P_+\g^bk_b\g^c\g^5\g^5\sigma+\g^a\g^5P_+\g^5\sigma\g^c\g^5\g^dk_d' \right) \nonumber
\\
&=\frac{\sigma}{2}\Tr\left( \g^a\g^5 \g^4\g^bk_b\g^c-\g^a\g^5\g^4\g^c\g^bk_b'\right) \nonumber
\\
&=\frac{\sigma}{2}\Tr\left(\g^a\g^5\g^b\g^c\g^4k_b-\g^a\g^5\g^c\g^b\g^4k_b'\right) \nonumber
\\
&=\frac{\sigma}{2}\Tr \left(-\g^a\g^b\g^c\g^4\g^5k_b+\g^a\g^c\g^b\g^4\g^5k_b'\right) \nonumber
\\
&=-\frac{\sigma}{2}\Tr\left[ \g^a\g^b\g^c\g^4\g^5\left(2k_b-q_b\right)\right] = 2i\sigma\epsilon^{abc}(2k_b-q_b).
\end{align}

Finally, it remains to compute the contribution from $S_{\text{f}}S_{\text{l}}$.
\begin{align}
&\quad \Tr\left[\g^a\g^5\left(\g^bk_b+\g^5\sigma\right)\g^c\g^5P_+\left(\g^dk'_d+\g^5\sigma\right)\right] \nonumber
\\
&=\Tr\left(\g^a\g^5\g^bk_b\g^c\g^5P_+\g^5\sigma+\g^a\g^5\g^5\sigma\g^c\g^5P_+\g^vk_d'\right)\nonumber
\\
&=\frac{\sigma}{2} \Tr\left(-\g^a\g^5\g^bk_b\g^c\g^4+\g^a\g^c\g^5\g^4\g^dk_d'\right) \nonumber
\\
&=\frac{\sigma}{2}\Tr\left[\g^a\g^b\g^c\g^4\g^5\left(2k_b-q_b\right)\right] = -2i\sigma\epsilon^{abc}(2k_b-q_b).
\end{align}
We note that this contribution is of similar type as from the other cross term, but with a difference in sign. These terms will, however, not cancel, which will be apparent in the following section. We also note that both the contributing terms have a term proportional to $k$, which is not the case for $\gamma^a$-coupling. This is because the $\gamma^5$-matrices introduce an additional relative sign between the corresponding terms.

%%%%%%%%%%%%%%%%%%%%%%%%%%%%%%%%%%%%%%%%%%%%%%%%

\subsection{Integral Calculations}
From the trace calculations, we see that the terms of interest in the propagator will look like,
\begin{align}
\Pi^{ac} = &-\frac{1}{2}\int \frac{d^3k}{(2\pi)^3}\int dx dx' \frac {e^{-|x-x'|k_3}}{k_3}\frac{e^{-|x-x'|k_3'}}{k_3'} \left[\frac{\mu\left(k'_3;x,x'\right)}{k'^2+\sigma^2}-\frac{\mu\left(k_3;x,x'\right)}{k^2+\sigma^2}\right]\frac{2i\sigma\left(2k_b-q_b\right)\epsilon^{abc}}{4},
\end{align}
where
\begin{equation}
k_3 := \sqrt{k^2+m_0^2+\sigma^2}; \quad k_3' := \sqrt{k'^2+m_0^2+\sigma^2}; \quad k' := k-q.
\end{equation}
Let us first focus on the integrand. Expanding $\frac{1}{k'^2+\sigma^2}$ around $q=0$, keeping at most linear terms in $q$ results in,
\begin{align}
&\quad \frac{e^{-\left|x-x'\right|\left(k_3+k_3'\right)}}{k_3k_3'} \left[\frac{\mu\left(k'_3;x,x'\right)}{k'^2+\sigma^2}-\frac{\mu\left(k_3;x,x'\right)}{k^2+\sigma^2}\right] \nonumber
\\
&= \frac{e^{-\left|x-x'\right|\left(k_3+k_3'\right)}}{k_3k_3'}\left\{\frac{1}{k^2+\sigma^2}\left[\mu\left(k_3';x,x'\right)-\mu\left(k_3;x,x'\right)\right]+\frac{\mu\left(k_3;x,x'\right)}{\left(k^2+\sigma^2\right)^2}2k\cdot q + \mathcal{O}\left(q^2\right)\right\}.
\end{align}
The difference becomes,
\begin{align}
\mu\left(k'_3;x,x'\right)-\mu\left(k_3;x,x'\right) &=2\frac{\left(k_3-k_3'\right)T_1}{k_3+k_3'}
\end{align}
with
\begin{equation}
T_1 := \frac{m_0\left(k_3+k_3'\right)\sgn\left(x-x'\right)}{2}\left[\tanh\left(m_0x\right)-\tanh\left(m_0x'\right)\right].
\end{equation}
Expanding this difference around $q=0$, again keeping at most linear terms in $q$, yields,
\begin{equation}
2\frac{\left(k_3-k_3'\right)T_1}{k_3+k_3'} = \frac{T_1 k\cdot q}{k_3^2} + \mathcal{O}\left(q^2\right).
\end{equation}
Introducing the variables
\begin{equation}
z = x-x'; \quad Z= x+x'
\end{equation}
allows us to write the integral on the following form
\begin{align}
\Pi^{ac} = -\frac{i\sigma\epsilon^{abc}}{4}\int\frac{d^3k}{\left(2\pi\right)^3}k_b\int dzdZ &\left\{ e^{-2|z|k_3}T_1k\cdot q\left[\frac{1}{k_3^4\left(k^2+\sigma^2\right)}-\frac{2}{k_3^2\left(k^2+\sigma^2\right)^2}\right]\right. \nonumber
\\
&-\left.e^{-2|z|k_3}T_2 2k\cdot q \frac{1}{k_3^2\left(k^2+\sigma^2\right)^2}\right\},
\end{align}
where
\begin{equation}
T_2:=m_0^2\left[1-\tanh\left(m_0x\right)\tanh\left(m_0x\right)\right].
\end{equation}
Using the integral relations in \oncite{mb},
\begin{align}
\frac{1}{2}\int dzdZe^{-2|z|k_3}T_1 &= \frac{m_0}{k_3},
\\
\frac{1}{2}\int dzdZe^{-2|z|k_3}T_2 &= \left[-\frac{m_0^2}{k_3^2}+2m_0\partial_{k_3}\psi\left(\frac{k_3}{m_0}\right)\right],
\end{align}
with
\begin{equation}
\partial_{k_3}\psi\left(\frac{k_3}{m_0}\right) := \frac{1}{m_0}\sum_{n=0}^{\infty}\frac{1}{\left(\frac{k_3}{m_0}+n\right)^2}
\end{equation}
being the trigamma function, we get
\begin{align}
\Pi^{ac} = -\frac{i\sigma \epsilon^{abc}}{2}\int \frac{d^3k}{\left(2\pi\right)^3}k_b &\left\{ m_0 k\cdot q\left[\frac{1}{\left(k^2+m_0^2+\sigma^2\right)^{\frac{5}{2}}\left(k^2+\sigma^2\right)}-\frac{2}{\left(k^2+m_0^2+\sigma^2\right)^{\frac{3}{2}}\left(k^2+\sigma^2\right)^2}\right] \right. \nonumber
\\
&+\left.\left[\frac{m_0^2}{k^2+m_0^2+\sigma^2}-2m_0\partial_{k_3}\psi\left(\frac{k_3}{m_0}\right)\right]\frac{2k\cdot q}{\left(k^2+m_0^2+\sigma^2\right)\left(k^2+\sigma^2\right)^2}\right\}.
\end{align}
Let us now investigate how to deal with this integral. We know that,
\begin{align}
\int \frac{d^3k}{\left(2\pi\right)^3}k_dq_ek_bg^{de}\epsilon^{abc} = q_eg^{de}g_{db}\epsilon^{abc}\frac{1}{3}\int \frac{d^3k}{\left(2\pi\right)^3}k^2 = q_b \epsilon^{abc}\frac{1}{3}\int \frac{d^3k}{\left(2\pi\right)^3}k^2.
\end{align}
By using this, the only dependence is in $k^2$, meaning that we can go to spherical coordinates and trivially perform the angular part. We end up with,
\begin{align} \label{eq:finint}
\Pi^{ac} = -\frac{i\sigma\epsilon^{abc}q_b}{6}\int_0^{\infty} \frac{dk}{2\pi^2}k^4&\left[\frac{m_0}{\left(k^2+m_0^2+\sigma^2\right)^{\frac{5}{2}}\left(k^2+\sigma^2\right)}-\frac{2m_0}{\left(k^2+m_0^2+\sigma^2\right)^{\frac{3}{2}}\left(k^2+\sigma^2\right)^2} \right. \nonumber
\\
&+\left.\frac{2m_0^2}{\left(k^2+m_0^2+\sigma^2\right)^{2}\left(k^2+\sigma^2\right)^2}-\frac{4m_0\partial_{k_3}\psi\left(\frac{k_3}{m_0}\right)}{\left(k^2+m_0^2+\sigma^2\right)\left(k^2+\sigma^2\right)^2}\right].
\end{align} 
By first investigating the behavior for small $\sigma$, we introduce the variable $K=\frac{k}{m_0}$, and note that Eq. \eqref{eq:finint} attains the form,
\begin{align} \label{eq:lastint}
\Pi^{ac} = -\frac{i\epsilon^{abc}q_b}{6}\frac{\sigma}{m_0} \int_0^{\infty} \frac{dK}{2\pi^2}K^4&\left[\frac{1}{\left(K^2+1+\frac{\sigma^2}{m_0^2}\right)^{\frac{5}{2}}\left(K^2+\frac{\sigma^2}{m_0^2}\right)}-\frac{2}{\left(K^2+1+\frac{\sigma^2}{m_0^2}\right)^{\frac{3}{2}}\left(K^2+\frac{\sigma^2}{m_0^2}\right)^2} \right. \nonumber
\\
&+ \left. \frac{2}{\left( K^2 + 1 +\frac{\sigma^2}{m_0^2}\right)^2\left(K^2 + \frac{\sigma^2}{m_0^2}\right)^2}-\frac{4\sum_{n=0}^{\infty} \left(\sqrt{K^2+1+\frac{\sigma^2}{m_0^2}}+n\right)^{-2}}{\left(K^2+1+\frac{\sigma^2}{m_0^2}\right)\left(K^2 + \frac{\sigma^2}{m_0^2}\right)^2}\right]. 
\end{align}
Because of the trigamma function, this integral is not analytically solvable. The numerical solution is graphically illustrated in Figure \ref{fig:hallcoef} (a) in the main text. The value of the derivative w.r.t. $\frac{\sigma}{m_0}$ at $\frac{\sigma}{m_0}=0$ is,
\begin{equation} \label{eq:lincscoef}
\Pi^{ac} \xrightarrow[\frac{\sigma}{m_0}\to 0]{}-\frac{\sigma}{m_0}i\epsilon^{abc}q_b\left(\frac{3\pi-10}{72\pi^2}-0.0503786\right).
\end{equation}
By re-labelling $b$ and $c$, we end up with,
\begin{equation}
\Pi^{ab} \xrightarrow[\frac{\sigma}{m_0}\to 0]{} -\frac{\sigma}{m_0} i\epsilon^{abc}q_c\cdot0.051188 .
\end{equation}

Considering the opposite limit, {\emph{i.e.}}, $\frac{m_0}{\sigma}\to 0$, we can introduce the variable $\tilde{K} = \frac{k}{\sigma}$, which yields,
\begin{align}
\Pi^{ac} = -\frac{i\epsilon^{abc}q_b}{12\pi^2}\frac{m_0}{\sigma} &\int_0^{\infty} d\tilde{K}\tilde{K}^4 \left[\frac{1}{\left(\tilde{K}^2 +\frac{m_0^2}{\sigma^2}+1\right)^{\frac{5}{2}}\left(\tilde{K}^2 +1\right)}-\frac{2}{\left(\tilde{K}^2+\frac{m_0^2}{\sigma^2}+1\right)^{\frac{3}{2}}\left(\tilde{K}^2+1\right)} \right. \nonumber
\\
& \left. +2\frac{m_0}{\sigma}\frac{1}{\left(\tilde{K}^2+\frac{m_0^2}{\sigma^2}+1\right)^2\left(\tilde{K}^2+1\right)^2}-4\frac{m_0}{\sigma} \frac{\sum_{n=0}^{\infty} \left(\sqrt{\tilde{K}^2+\frac{m_0^2}{\sigma^2}+1}+n\frac{m_0}{\sigma}\right)^{-2}}{\left(\tilde{K}^2+\frac{m_0^2}{\sigma^2}+1\right)\left(\tilde{K}^2+1\right)^2}\right]
\end{align}
which again has to be performed numerically. A graphical illustration of the result is given in Figure \ref{fig:hallcoef} (b).

However, for the value of the derivative w.r.t. $\frac{m_0}{\sigma}$ at $\frac{m_0}{\sigma}=0$, we can use the asymptotic expansion to simplify the trigamma function,
\begin{equation}
\partial_{k_3}\psi\left(\frac{k_3}{m_0}\right) = \frac{1}{k_3}.
\end{equation}
Using this, introducing the variable $\tilde{K} = \frac{k}{\sigma}$ and keeping only leading order terms in $\frac{m_0}{\sigma}$ gives us the following integral,
\begin{align}
&\quad \Pi^{ac}\xrightarrow[\frac{m_0}{\sigma}\to 0]{} \nonumber
\\
&-\frac{i\epsilon^{abc}q_b}{12\pi ^2}\frac{m_0}{\sigma} \int_0^{\infty}d\tilde{K}\tilde{K}^4\left[\frac{1}{\left(\tilde{K}^2+\frac{m_0^2}{\sigma^2}+1\right)^{\frac{5}{2}}\left(\tilde{K}^2+1\right)}-\frac{6}{\left(\tilde{K}^2+\frac{m_0^2}{\sigma^2}+1\right)^{\frac{3}{2}}\left(\tilde{K}^2+1\right)^2}\right]. \label{eq:smallmint}
\end{align}
As above, the value of the derivative at $\frac{m_0}{\sigma}=0$ can be evaluated by setting $\frac{m_0^2}{\sigma^2}=0$ in the integrand, which yields,
\begin{equation}
\Pi^{ab}\xrightarrow[\frac{m_0}{\sigma}\to 0]{}-\frac{m_0}{\sigma}\frac{\epsilon^{abc} q_c}{12\pi^2},
\end{equation}
which is consistent with computing the current from the triangular graph in Figure \ref{fig:diag}.

\end{widetext}
%%%%%%%%%%%%%%%%%%%%

\bibliography{TSC}

\begin{thebibliography}{19}%
\makeatletter
\providecommand \@ifxundefined [1]{%
 \@ifx{#1\undefined}
}%
\providecommand \@ifnum [1]{%
 \ifnum #1\expandafter \@firstoftwo
 \else \expandafter \@secondoftwo
 \fi
}%
\providecommand \@ifx [1]{%
 \ifx #1\expandafter \@firstoftwo
 \else \expandafter \@secondoftwo
 \fi
}%
\providecommand \natexlab [1]{#1}%
\providecommand \enquote  [1]{``#1''}%
\providecommand \bibnamefont  [1]{#1}%
\providecommand \bibfnamefont [1]{#1}%
\providecommand \citenamefont [1]{#1}%
\providecommand \href@noop [0]{\@secondoftwo}%
\providecommand \href [0]{\begingroup \@sanitize@url \@href}%
\providecommand \@href[1]{\@@startlink{#1}\@@href}%
\providecommand \@@href[1]{\endgroup#1\@@endlink}%
\providecommand \@sanitize@url [0]{\catcode `\\12\catcode `\$12\catcode
  `\&12\catcode `\#12\catcode `\^12\catcode `\_12\catcode `\%12\relax}%
\providecommand \@@startlink[1]{}%
\providecommand \@@endlink[0]{}%
\providecommand \url  [0]{\begingroup\@sanitize@url \@url }%
\providecommand \@url [1]{\endgroup\@href {#1}{\urlprefix }}%
\providecommand \urlprefix  [0]{URL }%
\providecommand \Eprint [0]{\href }%
\providecommand \doibase [0]{http://dx.doi.org/}%
\providecommand \selectlanguage [0]{\@gobble}%
\providecommand \bibinfo  [0]{\@secondoftwo}%
\providecommand \bibfield  [0]{\@secondoftwo}%
\providecommand \translation [1]{[#1]}%
\providecommand \BibitemOpen [0]{}%
\providecommand \bibitemStop [0]{}%
\providecommand \bibitemNoStop [0]{.\EOS\space}%
\providecommand \EOS [0]{\spacefactor3000\relax}%
\providecommand \BibitemShut  [1]{\csname bibitem#1\endcsname}%
\let\auto@bib@innerbib\@empty
%</preamble>
\bibitem [{\citenamefont {Schnyder}\ \emph {et~al.}(2008)\citenamefont
  {Schnyder}, \citenamefont {Ryu}, \citenamefont {Furusaki},\ and\
  \citenamefont {Ludwig}}]{PhysRevB.78.195125}%
  \BibitemOpen
  \bibfield  {author} {\bibinfo {author} {\bibfnamefont {A.~P.}\ \bibnamefont
  {Schnyder}}, \bibinfo {author} {\bibfnamefont {S.}~\bibnamefont {Ryu}},
  \bibinfo {author} {\bibfnamefont {A.}~\bibnamefont {Furusaki}}, \ and\
  \bibinfo {author} {\bibfnamefont {A.~W.~W.}\ \bibnamefont {Ludwig}},\ }\href
  {\doibase 10.1103/PhysRevB.78.195125} {\bibfield  {journal} {\bibinfo
  {journal} {Phys. Rev. B}\ }\textbf {\bibinfo {volume} {78}},\ \bibinfo
  {pages} {195125} (\bibinfo {year} {2008})}\BibitemShut {NoStop}%
\bibitem [{\citenamefont {Kitaev}(209)}]{kitaev2009}%
  \BibitemOpen
  \bibfield  {author} {\bibinfo {author} {\bibfnamefont {A.}~\bibnamefont
  {Kitaev}},\ }\href {https://doi.org/10.1063/1.3149495} {\bibfield  {journal}
  {\bibinfo  {journal} {Advances in Theoretical Physics: Landau Memorial
  Conference, edited by V. Lebedev and M. Feigelman}\ }\textbf {\bibinfo
  {volume} {AIP Conference Proceedings, 1134}},\ \bibinfo {pages} {22}
  (\bibinfo {year} {209})}\BibitemShut {NoStop}%
\bibitem [{\citenamefont {Ryu}\ \emph {et~al.}(2012)\citenamefont {Ryu},
  \citenamefont {Moore},\ and\ \citenamefont {Ludwig}}]{PhysRevB.85.045104}%
  \BibitemOpen
  \bibfield  {author} {\bibinfo {author} {\bibfnamefont {S.}~\bibnamefont
  {Ryu}}, \bibinfo {author} {\bibfnamefont {J.~E.}\ \bibnamefont {Moore}}, \
  and\ \bibinfo {author} {\bibfnamefont {A.~W.~W.}\ \bibnamefont {Ludwig}},\
  }\href {\doibase 10.1103/PhysRevB.85.045104} {\bibfield  {journal} {\bibinfo
  {journal} {Phys. Rev. B}\ }\textbf {\bibinfo {volume} {85}},\ \bibinfo
  {pages} {045104} (\bibinfo {year} {2012})}\BibitemShut {NoStop}%
\bibitem [{\citenamefont {Qi}\ \emph {et~al.}(2013)\citenamefont {Qi},
  \citenamefont {Witten},\ and\ \citenamefont {Zhang}}]{qwz}%
  \BibitemOpen
  \bibfield  {author} {\bibinfo {author} {\bibfnamefont {X.-L.}\ \bibnamefont
  {Qi}}, \bibinfo {author} {\bibfnamefont {E.}~\bibnamefont {Witten}}, \ and\
  \bibinfo {author} {\bibfnamefont {S.-C.}\ \bibnamefont {Zhang}},\ }\href
  {\doibase 10.1103/PhysRevB.87.134519} {\bibfield  {journal} {\bibinfo
  {journal} {Phys. Rev. B}\ }\textbf {\bibinfo {volume} {87}},\ \bibinfo
  {pages} {134519} (\bibinfo {year} {2013})}\BibitemShut {NoStop}%
\bibitem [{\citenamefont {Stone}\ and\ \citenamefont {Lopes}(2016)}]{sl}%
  \BibitemOpen
  \bibfield  {author} {\bibinfo {author} {\bibfnamefont {M.}~\bibnamefont
  {Stone}}\ and\ \bibinfo {author} {\bibfnamefont {P.~L. e.~S.}\ \bibnamefont
  {Lopes}},\ }\href {\doibase 10.1103/PhysRevB.93.174501} {\bibfield  {journal}
  {\bibinfo  {journal} {Phys. Rev. B}\ }\textbf {\bibinfo {volume} {93}},\
  \bibinfo {pages} {174501} (\bibinfo {year} {2016})}\BibitemShut {NoStop}%
\bibitem [{\citenamefont {Nogueira}\ \emph {et~al.}(2016)\citenamefont
  {Nogueira}, \citenamefont {Nussinov},\ and\ \citenamefont {van~den
  Brink}}]{PhysRevLett.117.167002}%
  \BibitemOpen
  \bibfield  {author} {\bibinfo {author} {\bibfnamefont {F.~S.}\ \bibnamefont
  {Nogueira}}, \bibinfo {author} {\bibfnamefont {Z.}~\bibnamefont {Nussinov}},
  \ and\ \bibinfo {author} {\bibfnamefont {J.}~\bibnamefont {van~den Brink}},\
  }\href {\doibase 10.1103/PhysRevLett.117.167002} {\bibfield  {journal}
  {\bibinfo  {journal} {Phys. Rev. Lett.}\ }\textbf {\bibinfo {volume} {117}},\
  \bibinfo {pages} {167002} (\bibinfo {year} {2016})}\BibitemShut {NoStop}%
\bibitem [{\citenamefont {Mulligan}\ and\ \citenamefont {Burnell}(2013)}]{mb}%
  \BibitemOpen
  \bibfield  {author} {\bibinfo {author} {\bibfnamefont {M.}~\bibnamefont
  {Mulligan}}\ and\ \bibinfo {author} {\bibfnamefont {F.~J.}\ \bibnamefont
  {Burnell}},\ }\href {\doibase 10.1103/PhysRevB.88.085104} {\bibfield
  {journal} {\bibinfo  {journal} {Phys. Rev. B}\ }\textbf {\bibinfo {volume}
  {88}},\ \bibinfo {pages} {085104} (\bibinfo {year} {2013})}\BibitemShut
  {NoStop}%
\bibitem [{\citenamefont {Srednicki}(2007)}]{srednickibook}%
  \BibitemOpen
  \bibfield  {author} {\bibinfo {author} {\bibfnamefont {M.}~\bibnamefont
  {Srednicki}},\ }\href@noop {} {\emph {\bibinfo {title} {Quantum Field
  Theory}}}\ (\bibinfo  {publisher} {Cambride University Press},\ \bibinfo
  {year} {2007})\BibitemShut {NoStop}%
\bibitem [{\citenamefont {Callan}\ and\ \citenamefont
  {Harvey}(1985)}]{callan1985anomalies}%
  \BibitemOpen
  \bibfield  {author} {\bibinfo {author} {\bibfnamefont {C.~G.}\ \bibnamefont
  {Callan}, \bibfnamefont {Jr}}\ and\ \bibinfo {author} {\bibfnamefont {J.~A.}\
  \bibnamefont {Harvey}},\ }\href {\doibase
  http://dx.doi.org/10.1016/0550-3213(85)90489-4} {\bibfield  {journal}
  {\bibinfo  {journal} {Nuclear Physics B}\ }\textbf {\bibinfo {volume}
  {250}},\ \bibinfo {pages} {427} (\bibinfo {year} {1985})}\BibitemShut
  {NoStop}%
\bibitem [{\citenamefont {Hansson}\ \emph {et~al.}(2004)\citenamefont
  {Hansson}, \citenamefont {Oganesyan},\ and\ \citenamefont
  {Sondhi}}]{HANSSON2004497}%
  \BibitemOpen
  \bibfield  {author} {\bibinfo {author} {\bibfnamefont {T.}~\bibnamefont
  {Hansson}}, \bibinfo {author} {\bibfnamefont {V.}~\bibnamefont {Oganesyan}},
  \ and\ \bibinfo {author} {\bibfnamefont {S.}~\bibnamefont {Sondhi}},\ }\href
  {\doibase https://doi.org/10.1016/j.aop.2004.05.006} {\bibfield  {journal}
  {\bibinfo  {journal} {Annals of Physics}\ }\textbf {\bibinfo {volume}
  {313}},\ \bibinfo {pages} {497} (\bibinfo {year} {2004})}\BibitemShut
  {NoStop}%
\bibitem [{\citenamefont {Kane}\ and\ \citenamefont
  {Mele}(2005)}]{PhysRevLett.95.226801}%
  \BibitemOpen
  \bibfield  {author} {\bibinfo {author} {\bibfnamefont {C.~L.}\ \bibnamefont
  {Kane}}\ and\ \bibinfo {author} {\bibfnamefont {E.~J.}\ \bibnamefont
  {Mele}},\ }\href {\doibase 10.1103/PhysRevLett.95.226801} {\bibfield
  {journal} {\bibinfo  {journal} {Phys. Rev. Lett.}\ }\textbf {\bibinfo
  {volume} {95}},\ \bibinfo {pages} {226801} (\bibinfo {year}
  {2005})}\BibitemShut {NoStop}%
\bibitem [{\citenamefont {Hasan}\ and\ \citenamefont
  {Kane}(2010)}]{RevModPhys.82.3045}%
  \BibitemOpen
  \bibfield  {author} {\bibinfo {author} {\bibfnamefont {M.~Z.}\ \bibnamefont
  {Hasan}}\ and\ \bibinfo {author} {\bibfnamefont {C.~L.}\ \bibnamefont
  {Kane}},\ }\href {\doibase 10.1103/RevModPhys.82.3045} {\bibfield  {journal}
  {\bibinfo  {journal} {Rev. Mod. Phys.}\ }\textbf {\bibinfo {volume} {82}},\
  \bibinfo {pages} {3045} (\bibinfo {year} {2010})}\BibitemShut {NoStop}%
\bibitem [{\citenamefont {Qi}\ and\ \citenamefont
  {Zhang}(2011)}]{RevModPhys.83.1057}%
  \BibitemOpen
  \bibfield  {author} {\bibinfo {author} {\bibfnamefont {X.-L.}\ \bibnamefont
  {Qi}}\ and\ \bibinfo {author} {\bibfnamefont {S.-C.}\ \bibnamefont {Zhang}},\
  }\href {\doibase 10.1103/RevModPhys.83.1057} {\bibfield  {journal} {\bibinfo
  {journal} {Rev. Mod. Phys.}\ }\textbf {\bibinfo {volume} {83}},\ \bibinfo
  {pages} {1057} (\bibinfo {year} {2011})}\BibitemShut {NoStop}%
\bibitem [{\citenamefont {Hsieh}\ \emph {et~al.}(2016)\citenamefont {Hsieh},
  \citenamefont {Cho},\ and\ \citenamefont {Ryu}}]{PhysRevB.93.075135}%
  \BibitemOpen
  \bibfield  {author} {\bibinfo {author} {\bibfnamefont {C.-T.}\ \bibnamefont
  {Hsieh}}, \bibinfo {author} {\bibfnamefont {G.~Y.}\ \bibnamefont {Cho}}, \
  and\ \bibinfo {author} {\bibfnamefont {S.}~\bibnamefont {Ryu}},\ }\href
  {\doibase 10.1103/PhysRevB.93.075135} {\bibfield  {journal} {\bibinfo
  {journal} {Phys. Rev. B}\ }\textbf {\bibinfo {volume} {93}},\ \bibinfo
  {pages} {075135} (\bibinfo {year} {2016})}\BibitemShut {NoStop}%
\bibitem [{\citenamefont {V\"ayrynen}\ and\ \citenamefont
  {Volovik}(2011)}]{vv}%
  \BibitemOpen
  \bibfield  {author} {\bibinfo {author} {\bibfnamefont {J.~I.}\ \bibnamefont
  {V\"ayrynen}}\ and\ \bibinfo {author} {\bibfnamefont {G.~E.}\ \bibnamefont
  {Volovik}},\ }\href {\doibase 10.1134/S0021364011060129} {\bibfield
  {journal} {\bibinfo  {journal} {JETP Letters}\ }\textbf {\bibinfo {volume}
  {93}} (\bibinfo {year} {2011}),\ 10.1134/S0021364011060129}\BibitemShut
  {NoStop}%
\bibitem [{\citenamefont {Nielsen}\ and\ \citenamefont
  {Ninomiya}(1983)}]{nielsen1983}%
  \BibitemOpen
  \bibfield  {author} {\bibinfo {author} {\bibfnamefont {H.}~\bibnamefont
  {Nielsen}}\ and\ \bibinfo {author} {\bibfnamefont {M.}~\bibnamefont
  {Ninomiya}},\ }\href {\doibase https://doi.org/10.1016/0370-2693(83)91529-0}
  {\bibfield  {journal} {\bibinfo  {journal} {Physics Letters B}\ }\textbf
  {\bibinfo {volume} {130}},\ \bibinfo {pages} {389 } (\bibinfo {year}
  {1983})}\BibitemShut {NoStop}%
\bibitem [{\citenamefont {Goldstone}\ and\ \citenamefont
  {Wilczek}(1981)}]{PhysRevLett.47.986}%
  \BibitemOpen
  \bibfield  {author} {\bibinfo {author} {\bibfnamefont {J.}~\bibnamefont
  {Goldstone}}\ and\ \bibinfo {author} {\bibfnamefont {F.}~\bibnamefont
  {Wilczek}},\ }\href {\doibase 10.1103/PhysRevLett.47.986} {\bibfield
  {journal} {\bibinfo  {journal} {Phys. Rev. Lett.}\ }\textbf {\bibinfo
  {volume} {47}},\ \bibinfo {pages} {986} (\bibinfo {year} {1981})}\BibitemShut
  {NoStop}%
\bibitem [{\citenamefont {Sp{\aa}nsl\"att}\ \emph {et~al.}(2015)\citenamefont
  {Sp{\aa}nsl\"att}, \citenamefont {Ardonne}, \citenamefont {Budich},\ and\
  \citenamefont {Hansson}}]{Spanslatt2015}%
  \BibitemOpen
  \bibfield  {author} {\bibinfo {author} {\bibfnamefont {C.}~\bibnamefont
  {Sp{\aa}nsl\"att}}, \bibinfo {author} {\bibfnamefont {E.}~\bibnamefont
  {Ardonne}}, \bibinfo {author} {\bibfnamefont {J.~C.}\ \bibnamefont {Budich}},
  \ and\ \bibinfo {author} {\bibfnamefont {T.~H.}\ \bibnamefont {Hansson}},\
  }\href {\doibase 10.1088/0953-8984/27/40/405701} {\bibfield  {journal}
  {\bibinfo  {journal} {Journal of Physics: Condensed Matter}\ }\textbf
  {\bibinfo {volume} {27}},\ \bibinfo {pages} {405701} (\bibinfo {year}
  {2015})}\BibitemShut {NoStop}%
\bibitem [{\citenamefont {M.~Stone}\ and\ \citenamefont
  {Goldbart}(2009)}]{mikebook}%
  \BibitemOpen
  \bibfield  {author} {\bibinfo {author} {\bibfnamefont {M.}~\bibnamefont
  {M.~Stone}}\ and\ \bibinfo {author} {\bibfnamefont {P.}~\bibnamefont
  {Goldbart}},\ }\href@noop {} {\emph {\bibinfo {title} {Mathematics for
  Physics}}}\ (\bibinfo  {publisher} {Cambridge University Press},\ \bibinfo
  {year} {2009})\BibitemShut {NoStop}%
\end{thebibliography}%



\end{document}